%% file: main.tex
\documentclass{article}
\usepackage[margin=1in]{geometry}
\input{packages}

\title{Technical Report: Match-reference regular expressions and lenses}
\author{Jeanne-Marie Musca \and Anders Miltner \and Kathleen Fisher \and David Walker}
\date{ }
\input{macros/general_formatting}
\input{macros/match-reference}

\input{macros/mras}

\input{macros/scoped_regex}

\input{macros/mrlens}

\newif\ifdraft\drafttrue      

\newif\ifappendix\appendixtrue   

\input{macros/colors}

\newcommand{\ename}{match-reference}
\newcommand{\upename}{Match-Reference}
\newcommand{\initename}{MR}
\newcommand{\ras}{RAS}
\newcommand{\xgenop}{\relax\ifmmode\mathop{\text{\normalfont \textsf{op}}}\else\textsf{op}\fi}

\theoremstyle{definition}
\newtheorem{example}{Example}[section]

\newtheorem{definition}{Definition}

\begin{document}

\maketitle

\section*{Abstract}

A \emph{lens} is a single program that specifies two data transformations at once: one transformation converts data from source format to target format and a second transformation inverts the process.  Over the past decade, researchers have developed many different kinds of lenses with different properties.  One class of such languages operate over regular languages.  In other words, these lenses convert strings drawn from one regular language to strings drawn from another regular language (and back again). In this paper, we define a more powerful language of lenses, which we call \emph{match-reference lenses}, that is capable of translating between non-regular formats that contain repeated substrings, which is a primitive form of \emph{dependency}.  To define the non-regular formats themselves, we develop a new language, \emph{match-reference regular expressions}, which are regular expressions that can bind variables to substrings and use those substrings repeatedly.  These match-reference regular expressions are closely related to the familiar ``back-references" that can be found in traditional regular expression packages, but are redesigned to adhere to conventional programming language lexical scoping conventions and to interact smoothly with lens language infrastructure.  We define the semantics of match-reference regular expressions and match-reference lenses. We also define a new kind of automaton, the \emph{match-reference regex automaton system} (MRRAS), for deciding string membership in the language match-reference regular expressions.  We illustrate our definitions with a variety of examples.

\section{Introduction}

Many programming tasks require transforming data between pairs of formats, such as: serializing and de-serializing data when storing or transmitting it, using a suite of programs where each program requires the data to be in its own ad-hoc representation, or parsing input and later pretty-printing it. As a result, it is often necessary to write pairs of translations -- a function and its inverse -- in order to manipulate data across pairs of representations. However, it can be difficult to keep these pairs of functions coordinated, especially when a format description is complex or requires updating.

\begin{figure}[t]
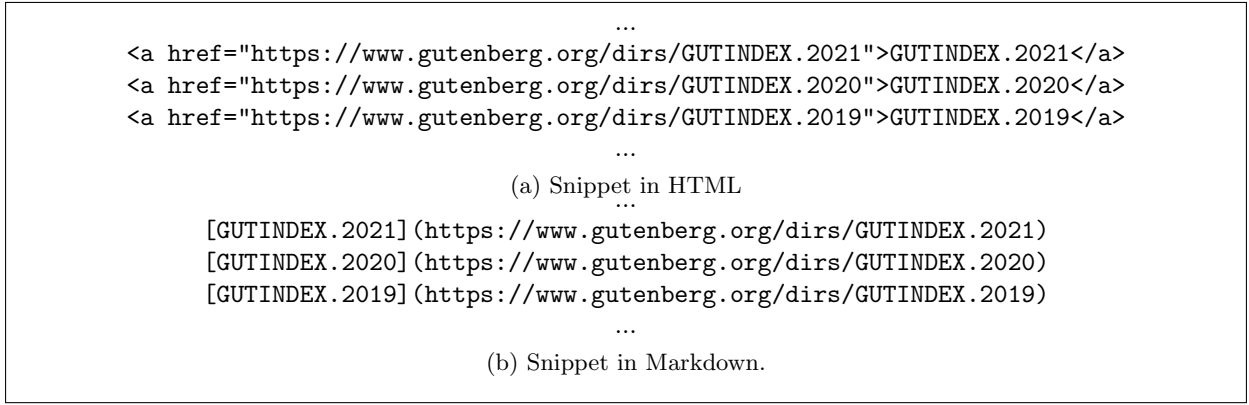

\begin{framed}
\begin{subfigure}{\textwidth}
    \centering
    ...

    \texttt{<a href="https://www.gutenberg.org/dirs/GUTINDEX.2021">GUTINDEX.2021</a>}

    \texttt{<a href="https://www.gutenberg.org/dirs/GUTINDEX.2020">GUTINDEX.2020</a>}

    \texttt{<a href="https://www.gutenberg.org/dirs/GUTINDEX.2019">GUTINDEX.2019</a>}

    ...

    \caption{Snippet in HTML}
    \label{fig:gut-ex}

\end{subfigure}
\begin{subfigure}{\textwidth}
    \centering
    ...

    \texttt{[GUTINDEX.2021](https://www.gutenberg.org/dirs/GUTINDEX.2021)}

    \texttt{[GUTINDEX.2020](https://www.gutenberg.org/dirs/GUTINDEX.2020)}

    \texttt{[GUTINDEX.2019](https://www.gutenberg.org/dirs/GUTINDEX.2019)}

    ...
    \caption{Snippet in Markdown.}
    \label{fig:gut-ex-md}
\end{subfigure}
\end{framed}
\caption{Simplified snippets of a Project Gutenberg webpage with hyperlinks to downloadable files \cite{PGLinks}}
\label{fig:gut-ex-both}
\end{figure}

To simplify the process of writing bidirectional transformations, Foster et al. introduce a domain-specific language that represents both a translation and its inverse  with a single expression: a \emph{lens} \cite{Foster2009, Foster2007}. Essentially, lenses encapsulate two functions: \lget\ and \lput. The \lget\ function transforms data from the source to the target format, while \lput\ does the inverse. Take the list of hyperlinks in \autoref{fig:gut-ex-both}: they are represented in both HTML and Markdown. In the lens DSL, we could write a lens, \xl, that translates the hyperlinks between HTML and Markdown. The \lget\ function for \xl\ would take the hyperlinks in HTML as input, and output them in Markdown, and \xl's \lput\ function would take the entries in Markdown and translate them into HTML.

At the core of the lens DSL is a set of bijective lens constructors that provide strong correctness guarantees but have limited expressiveness. They only translate between formats that can be represented by regular expressions. Many common formats are non-regular, however, with internal dependencies that come in various forms, such as length fields, matching xml tags, or the repeated filename, \texttt{GUTINDEX.\#\#\#\#}, that appears in the hyperlinks given in \autoref{fig:gut-ex-both}. Prior work augments the core lenses with constructors that increase the expressiveness of the DSL, but at the cost of providing weaker correctness guarantees. Our contribution, \emph{match-reference bijective lenses}, is an extension of the core lenses that can translate formats that have a foundational type of internal dependency --matching sub-strings-- and still provide strong correctness guarantees.

The correctness guarantees of the lens DSL are given in the context of a type system, where a lens' type is a pair of regular expressions that describes the two formats between which the lens translates. So, in order to maintain the correctness guarantees for the match-reference lens extension, we also extend the type system to allow for a richer description of the formats. Our second contribution, therefore, is an extension to regular expressions --\emph{match-reference regular expressions}-- which represents languages with matching sub-strings.

We have developed match-reference regular expressions not only as types for match-reference lenses, but also as a theoretically well-defined version of PCRE (perl-compatible regular expressions) back-references \cite{PCREman}. 
In their own right, regular expressions serve as the basis for recognizing and parsing all kinds of data. In practice, modern scripting languages augment regular expressions with non-regular features. PCRE back-references are a common extension that allow users to give a name to a string and then refer to that string repeatedly. PCRE back-references, however, are notoriously difficult to characterize theoretically \cite{Campeanu2009b,Schmid2012}, and so, in this context, we build on work that aims to craft a formalism that is as expressive as back-references while being theoretically robust.

This report is structured as follows:
\begin{itemize}
    \item \autoref{sec:background} provides background on core regular expressions and typed bijective lenses. It establishes the notation that we use in the rest of the report, shows how to write a basic lens for the example in \autoref{fig:gut-ex-both}, and gives the type for that lens.
    
    \item \autoref{sec:mre} gives the specification for match-reference regular expressions and compares them to PCRE back-references.
    
    \item \autoref{sec:mre-lens} characterizes match-reference typed bijective lenses, including a detailed specification of their type system.
    
    \item \autoref{sec:mre-lens-impl} gives the basis for implementing match-reference lenses in the form of a big-step semantics for the lenses, and a machine for deciding membership in the language of match-reference regexes.
\end{itemize}

\section{Context: Core regular expressions and bijective lenses 101}
\label{sec:background}

To establish notation and set the stage for our work, we briefly review regular expressions (REs) and bijective lenses. As a running example,
consider simplified fragments of a Project Gutenberg web page given in \autoref{fig:gut-ex-both}, in both HTML and Markdown. These fragments list download links organized by year. The HTML code \xlit{<a href="}\textbf{url}\xlit{">}\textbf{text}\xlit{</a>} displays in HTML viewers as the word \textbf{text};  clicking on \textbf{text} navigates to the URL \textbf{url}. The Markdown code \xlit{[}\textbf{text}\xlit{]}\xlit{(}\textbf{url}\xlit{)} when interpreted has the same behavior as that of the HTML code.

In what follows, we represent the formats of the snippets in \autoref{fig:gut-ex-both} with core regular expressions using the notation we adopt, and write a bijective lens to translate between these two formats. We give the type of the resulting lens using the regular expressions that represent the HTML and Markdown formats.

\subsection{Representing formats with regular expressions}

We use the following syntax for regular expressions:
\begin{center}
    \ginit{\mathR}
    \rconst{c}
    \gsep\ \rstar{\mathR}
    \gsep\ \rand{\mathR}{\mathR}
    \gsep\ \ralt{\mathR}{\mathR}
\end{center}
 A regular expression \mathR\ is a constant ($c$), or formed via iteration (\rstar{\mathR}), concatenation (\rand{\mathR}{\mathR}), or alternation (\ralt{\mathR}{\mathR}).
To increase legibility,
we adopt some standard conventions: we juxtapose regular expressions to denote concatenation, writing \xlit{</a>} instead of the more heavyweight
\rand{\xlit{<}}{\rand{\xlit{/}}{\rand{\xlit{a}}{\xlit{>}}}}, and we abbreviate the regular expression \ralt{\ralt{\ralt{\xlit{a}}{\ralt{\ldots}{\xlit{z}}}}{\ralt{\xlit{A}}{\ralt{\ldots}{\xlit{Z}}}}}{\ralt{\xlit{0}}{\ralt{\ldots}{\xlit{9}}}} by the more concise character class notation [\xlit{a}-\xlit{zA}-\xlit{Z0}-\xlit{9}]. We also adopt the convention of naming regular expressions for re-use in larger expression.

Using these conventions, we can easily represent the HTML and Markdown formats of the Project Gutenberg snippets given in \autoref{fig:gut-ex-both} using regular expressions.
First, we define a regular expression for representing the urls on the webpage:
\begin{center}
    \xrdef{pg\_url} := \xlit{https://www.gutenberg.org/dirs/GUTINDEX.}[\xlit{0}-\xlit{9}][\xlit{0}-\xlit{9}][\xlit{0}-\xlit{9}][\xlit{0}-\xlit{9}]
\end{center}
Next, we define a regular expression that represents the displayed file name:
\begin{center}
    \xrdef{pg\_fname} := \xlit{GUTINDEX.}[\xlit{0}-\xlit{9}][\xlit{0}-\xlit{9}][\xlit{0}-\xlit{9}][\xlit{0}-\xlit{9}]
\end{center}
With these two definitions, we now give regular expressions for a single line of the HTML and Markdown formats:
\begin{align*}
    \xrdef{pg\_html\_line} &:= \rand{\rand{\xlit{<a href="}}{\rand{\xrdef{pg\_url}}{\xlit{">}}}}
                              {\rand{\xrdef{pg\_fname}}{\xlit{</a>}}}
    \\
    \xrdef{pg\_md\_line} &:= \rand{\rand{\xlit{[}}{\rand{\xrdef{pg\_fname}}{\xlit{](}}}}
                                  {\rand{\xrdef{pg\_url}}{\xlit{)}}}
\end{align*}
 and star those expressions to represent a list of hyperlinks:
\begin{align*}
    \xrdef{pg\_html} &:= \rstar{\xrdef{pg\_html\_line} }
    \\
    \xrdef{pg\_md} &:= \rstar{\xrdef{pg\_md\_line}}
\end{align*}

\newcommand{\sxconst}[2]{\ensuremath{(#1, #2)}}
\newcommand{\sxiter}[1]{\ensuremath{{#1}^{*}}}
\newcommand{\sxconcat}[2]{\ensuremath{#1 . #2}}
\newcommand{\sxswap}[2]{\ensuremath{#1 \sim #2}}
\begin{figure}
\begin{subfigure}{\textwidth}
    \begin{align*}
\xrdef{pg\_url} &:= \xlit{https://www.gutenberg.org/dirs/GUTINDEX.}
        \text{[\xlit{0}-\xlit{9}][\xlit{0}-\xlit{9}][\xlit{0}-\xlit{9}][\xlit{0}-\xlit{9}]} \\
\xrdef{pg\_fname} &:= 
    \xlit{GUTINDEX.}
    \text{[\xlit{0}-\xlit{9}][\xlit{0}-\xlit{9}][\xlit{0}-\xlit{9}][\xlit{0}-\xlit{9}]} \\
  \xrdef{pg\_html\_line} &:= \rand{\rand{\xlit{<a href="}}{\rand{\xrdef{pg\_url}}{\xlit{">}}}}
                              {\rand{\xrdef{pg\_fname}}{\xlit{</a>}}}
    \\
    \xrdef{pg\_md\_line} &:= \rand{\rand{\xlit{[}}{\rand{\xrdef{pg\_fname}}{\xlit{](}}}}
                                  {\rand{\xrdef{pg\_url}}{\xlit{)}}} \\
\xrdef{pg\_html}  &:= 
    \rstar{\xrdef{pg\_html\_line}} \\
\xrdef{pg\_md} &:= \rstar{ \xrdef{pg\_md\_line}}    
    \end{align*}
    \caption{Regular expressions that represent the Project Gutenberg hyperlinks}
    \label{fig:pg-regex-summary}
\end{subfigure}
\begin{subfigure}{\textwidth}
\begin{align*}
\xldef{pg\_url\_map} & :=
  \cunderline{blue}{\xconst{\xlit{<a href="}}{\xlit{(}}}
  \ . \ \cunderline{red}{\xid{\xrdef{pg\_url}}}
  \ . \ \cunderline{teal}{\xconst{\xlit{">}}{\xlit{)}}} \\
\xldef{pg\_fname\_map} & :=
  \cunderline{cyan}{\xconst{ \ }{\xlit{[}}}
  \ . \ \cunderline{orange}{\xid{\xrdef{pg\_fname}}}
  \ . \ \cunderline{green}{\xconst{\xlit{</a>}}{\xlit{]}}} \\
    \xldef{pg\_line\_map} & :=
    \sxswap{{\cunderline{lightgray}{\xldef{pg\_url\_map}}}}{{\cunderline{black}{\xldef{pg\_fname\_map}}}} \\         
\xldef{pg\_map} &:= 
    \xiter{\xldef{pg\_line\_map}}
\end{align*}
\begin{center}%
    \xlit{\smaller \wcbox{lightgray}{\cbox{blue}{<a href="}\cbox{red}{https://www.gutenberg.org/dirs/GUTINDEX.2021}\cbox{teal}{">}}\wcbox{black}{\cbox{cyan}{}\cbox{orange}{GUTINDEX.2021}\cbox{green}{</a>}}}
    
    \gpt
        
        \xlit{\smaller \wcbox{black}{\cbox{cyan}{[}\cbox{orange}{GUTINDEX.2021}\cbox{green}{]}}\wcbox{lightgray}{\cbox{blue}{(}\cbox{red}{https://www.gutenberg.org/dirs/GUTINDEX.2021}\cbox{teal}{)}}}
\end{center}
\caption{Bijective lenses that translate the Project Gutenberg hyperlinks between HTML and Markdown with an example translation. Elements of the lens definitions are underlined with a color. Correspondingly colored boxes are put around the parts of the HTML and Markdown hyperlinks that would be translated by that lens.}
\label{fig:pg-lens-summary}
\end{subfigure}
\caption{Regular expressions and bijective lenses for the Project Gutenberg hyperlinks in \autoref{fig:gut-ex-both}}
\label{fig:pg-summary-both}
\end{figure}

\subsection{Writing a basic bijective lens}
\label{ssec:basic-lens-ex}

A lens translates data back and forth between two formats, such as HTML and Markdown, and keeps the data in sync across the formats. In essence, lenses bundle together two functions \lget\ and \lput: \lget\ transforms data from the source to the target format, while \lput\ does the inverse. \emph{Bijective} lenses, as their name indicates, are lenses  whose \lget\ and \lput\ functions are bijections. In this work, we adopt a set of bijective lens constructors introduced in a lens DSL by Foster et al. \cite{Foster2009,Foster2007}, with the following syntax:
\begin{center}
    \ginit{l}
        \xconst{\s[1]}{\s[2]}
        \gsep\ \xid{\mathR}
        \gsep\ \xiter{\xl}
        \gsep\ \xconcat{\xl[1]}{\xl[2]}
        \gsep\ \xswap{\xl[1]}{\xl[2]}
        \gsep\ \xcomp{\xl[1]}{\xl[2]}
        \gsep\ \xlor{\xl[1]}{\xl[2]}
\end{center}

The base lenses are the \emph{constant} lens, \xconst{\s[1]}{\s[2]}, which translates between two strings and the \emph{identity} lens, \xid{\mathR}, which applies the identity function to strings in the language of \mathR. Lenses can also be formed via the iteration of a lens, \xiter{\xl}, the concatenation of two lenses, \xconcat{\xl[1]}{\xl[2]}, swapping the output of two lenses, \xswap{\xl[1]}{\xl[2]}, the composition of two lenses \xcomp{\xl[1]}{\xl[2]} or the alternation of two lenses, \xlor{\xl[1]}{\xl[2]}.

With these lens constructors, we can build a bijective lens, \xldef{pg\_map}, whose source is a list of Project Gutenberg hyperlinks in HTML and  whose target is the list of hyperlinks in Markdown. In other words, \xldef{pg\_map}'s \lget\ function converts the hyperlinks on the Project Gutenberg page from HTML to Markdown, and \xldef{pg\_map}'s \lput\ function converts the hyperlinks from Markdown to HTML. 
To give an intuition for how the lens constructors work, we walk through the process of writing \xldef{pg\_map}.

We start by defining a lens, \xldef{pg\_url\_map}, that translates between the URL component of the two formats.
Both formats contain a string matching the regular expression \xrdef{pg\_url}, but the formats have different constant strings that surround said URLs: the \xlit{<a href="} ... \xlit{">} scaffolding for HTML and the parentheses for Markdown. The constant lens \xconst{\s[1]}{\s[2]} serves to translate the scaffolding across the formats. For instance, the lens \xconst{\xlit{<a href="}}{\xlit{(} } transforms the string \xlit{<a href="} to the string \xlit{(} (and back again). To keep the same URL in both the HTML and Markdown versions, we use the identity lens \xid{\mathR}; its \lget\ and \lput\ functions perform the identity transformation when provided a string in \mathR. In our case, \mathR\ is \xrdef{pg\_url}. Finally, we use \emph{concatenation} lenses (\xconcat{\xl[1]}{\xl[2]}), to combine the results of applying the constant and identity lenses to parts of the URL. The concatenation lens takes two lenses as arguments, applies each of the lenses to parts of the incoming string, and concatenates the results. In this example, we write the concatenation lens as \xl[1] . \xl[2], for legibility. The result is \xldef{pg\_url\_map}, given here, with an example translation (the colored boxes highlight the result of applying sublenses to subsections of the strings).

\begin{center}
\xldef{pg\_url\_map} :=
  \cunderline{blue}{\xconst{\xlit{<a href="}}{\xlit{(}}}
  \ . \cunderline{red}{\xid{\xrdef{pg\_url}}}
  \ . \cunderline{teal}{\xconst{\xlit{">}}{\xlit{)}}}

\smallskip  
 {\smaller \xlit{\cbox{blue}{<a href="}\cbox{red}{https://www.gutenberg.org/dirs/GUTINDEX.2021}\cbox{teal}{">}} \gpt\ \xlit{\cbox{blue}{(}\cbox{red}{https://www.gutenberg.org/dirs/GUTINDEX.2021}\cbox{teal}{)}}}
\end{center}

We similarly define, \xldef{pg\_fname\_map}, to transform the displayed file name:
\begin{center}
\xldef{pg\_fname\_map} :=
  \cunderline{cyan}{\xconst{ \ }{\xlit{[}}}
  \ . \cunderline{orange}{\xid{\xrdef{pg\_fname}}}
  \ . \cunderline{green}{\xconst{\xlit{</a>}}{\xlit{]}}}
  
  \smallskip
    
   {\smaller \xlit{\cbox{cyan}{ }\cbox{orange}{GUTINDEX.2021}\cbox{green}{</a>}} \gpt\ \xlit{\cbox{cyan}{[}\cbox{orange}{GUTINDEX.2021}\cbox{green}{]}}}
   
\end{center}
 The first constant lens in \xldef{pg\_fname\_map}, \xconst{ \ }{\xlit{[}}, translates between the empty string and \xlit{[}. 
 Again, constant lenses translate the scaffolding and an identity lens is used to keep the displayed name unchanged across the translation, provided that the name is correctly formatted, and in the language of \xrdef{pg\_fname}. 

Next, we want to combine the result of translating the URL and displayed name of a single entry. In this case, the concatenation lens does not serve us well, because the HTML and Markdown formats swap the order of elements: HTML has the URL first while Markdown puts the displayed filename first. So, we use the \emph{swap} lens (\xswap{\xl[1]}{\xl[2]}) abbreviated as \sxswap{\xl[1]}{\xl[2]} to achieve this transformation:
\begin{center}
    \xldef{pg\_line\_map} :=
    \sxswap{{\cunderline{lightgray}{\xldef{pg\_url\_map}}}}{{\cunderline{black}{\xldef{pg\_fname\_map}}}}

\medskip
    \xlit{\smaller \cbox{lightgray}{<a href="https://www.gutenberg.org/dirs/GUTINDEX.2021">}\cbox{black}{GUTINDEX.2021</a>}}
    
    \gpt
        
        \xlit{\smaller \cbox{black}{[GUTINDEX.2021]}\cbox{lightgray}{(https://www.gutenberg.org/dirs/GUTINDEX.2021)}}
\end{center}
The swap lens performs a similar transformation to the concatenation lens, but swaps the order in the process.
 So, \xldef{pg\_line\_map}'s \lget\ function first translates the URL from HTML to Markdown and then translates the displayed name from HTML to Markdown. Finally, the lens' \lget\ function swaps the two translated parts and concatenates those parts. In contrast, \xldef{pg\_line\_map}'s \lput\ function first translates the \emph{displayed name} from Markdown to HTML, and then translates the URL from Markdown to HTML. Similarly to the lens' \lget\ function, though, the \lput\ function then swaps the translated parts before concatenating them.

Finally, we write the lens that will transform a list of hyperlinks between HTML and Markdown.
\begin{center}
    \xldef{pg\_map} := \xiter{\xldef{pg\_line\_map}}
\end{center}
We use the \emph{iteration} lens, \xiter{\xl}, to iterate the transformation that \xl\ performs across the list of hyperlinks.

\subsection{Bijective lens typing}

Undisciplined use of the lens combinators can result in programs that either fail to process data in the desired format or that corrupt data as it is translated back and forth between the two formats. To avoid such problems, Foster et al. have defined a type system for bijective lenses based on regular expressions. The basic judgment \xl\ : \ltype{\mathR[1]}{\mathR[2]} stipulates that the lens \xl\ translates strings in the language of \mathR[1] to the language of \mathR[2] and vice versa.

The type system for the core bijective lenses is given in \autoref{fig:core-ltype}. The rules include side conditions that ensure that a lens type contains \emph{strongly unambiguous} regexes: regexes that uniquely parse each string in their language. In particular, when operators are modified with exclamation marks (\emath{*^!}, \emath{\cdot^!}, and \emath{+^!}), this means that they are unambiguous operations. This unambiguity in the lens types ensures that the result of applying a lens to a string is deterministic, which is essential for the strong correctness guarantees of well-typed bijective lenses.

Within the type system, the lenses defined above have the following types:
\begin{align*}
    \xldef{pg\_url\_map} & :  \ltype{\rand{\xlit{<a href="}}{\rand{\xrdef{pg\_url}}{\xlit{">}}}}
                                 {\rand{\xlit{(}}{\rand{\xrdef{pg\_url}}{\xlit{)}}}} \\
    \xldef{pg\_fname\_map}  &  : \ltype{\rand{\xrdef{pg\_fname}}{\xlit{</a>}} }
                                   {\rand{\xlit{[}}{\rand{\xrdef{pg\_fname}}{\xlit{]}}}} \\
\xldef{pg\_line\_map} & : \ltype{\rand{\rand{\xlit{<a href="}}{\rand{\xrdef{pg\_url}}{\xlit{">}}}}{\rand{\xrdef{pg\_fname}}{\xlit{</a>}}}}
                                        {\rand{\rand{\xlit{[}}{\rand{\xrdef{pg\_fname}}{\xlit{]}}}}{\rand{\xlit{(}}{\rand{\xrdef{pg\_url}}{\xlit{)}}}}} \\
\xldef{pg\_map} &: \ltype{\xrdef{pg\_html}}{\xrdef{pg\_md}}
\end{align*}
Note that we expand \xrdef{pg\_html\_line} and \xrdef{pg\_md\_line} in the type for \xldef{pg\_line\_map} to show that the lens does indeed swap the displayed text and underlying URL components when translating the hyperlinks between HTML and Markdown.

When a bijective lense is well-typed (\xl\ : \ltype{\mathR[1]}{\mathR[2]}), it provably has the following properties:

\begin{itemize}
    \item \xl.get : \mathL(\mathR[1]) $\rightarrow$ \mathL(\mathR[2])
    \item \xl.put : \mathL(\mathR[2]) $\rightarrow$ \mathL(\mathR[1])
    \item \xl.put(\xl.get(\s)) == \s \qquad \qquad \textsc{(GetPut)}
    \item \xl.get(\xl.put(\s)) == \s \qquad \qquad \textsc{(PutGet)}
\end{itemize}
The last two equations, \textsc{GetPut} and \textsc{PutGet}, are known as the round-tripping laws. \textsc{GetPut} implies that when \xl\ translates from source to target and back, the data remains unchanged. Likewise, \textsc{PutGet} implies that when one starts at the target, translates to the source, and back to the target, the data is again unchanged.

So because \xldef{pg\_map} is well-typed, we know that when we translate the Project Gutenberg snippet in \autoref{fig:gut-ex} from HTML to Markdown, we will get valid Markdown code that represents the same hyperlinks as those encoded in the HTML, and likewise when we use the lens to translate the snippet in \autoref{fig:gut-ex-md} from Markdown to HTML, we know that the translation will have the results we expect.
\begin{figure}
\begin{mathpar}
    \inferrule{ }{\xconst{\s[1]}{\s[2]} : \ltype{\s[1]}{\s[2]}}
    
    \inferrule{\mathR[] \text{ is strongly unambiguous}}{\xid{\mathR} : \ltype{\mathR}{\mathR}}
    
    \inferrule{\xl : \ltype{\mathR[1]}{\mathR[2]}
         \and \mathR[1]^{*!} \and \mathR[2]^{*!}}
         {\xl : \ltype{\rstar{\mathR[1]}}{\rstar{\mathR[2]}}}
     
    \inferrule{\xl[1] : \ltype{\mathR[11]}{\mathR[12]} 
                \and \xl[2] : \ltype{\mathR[21]}{\mathR[22]} 
                \\\\ \mathR[11] \cdot^! \mathR[21]
                \and \mathR[12] \cdot^! \mathR[22]
                }
                {\xconcat{\xl[1]}{\xl[2]} : \ltype{\rand{\mathR[11]}{\mathR[21]}}{\rand{\mathR[12]}{\mathR[22]}}}
                
    \inferrule{\xl[1] : \ltype{\mathR[11]}{\mathR[12]} 
                \and \xl[2] : \ltype{\mathR[21]}{\mathR[22]} 
                \\\\ \mathR[11] \cdot^! \mathR[22]
                \and \mathR[12] \cdot^! \mathR[21]
                }
                {\xswap{\xl[1]}{\xl[2]} : \ltype{\rand{\mathR[11]}{\mathR[22]}}{\rand{\mathR[12]}{\mathR[21]}}}
                
    \inferrule{\xl[1] : \ltype{\mathR[1]}{\mathR[2]} 
                \and \xl[2] : \ltype{\mathR[2]}{\mathR[3]}}
                {\xcomp{\xl[1]}{\xl[2]}: \ltype{\mathR[1]}{\mathR[2]}}
                
    \inferrule{\xl[1] : \ltype{\mathR[11]}{\mathR[12]} 
                \and \xl[2] : \ltype{\mathR[21]}{\mathR[22]} 
                \\\\ \mathR[11] +^! \mathR[21]
                \and \mathR[12] +^! \mathR[22]}
                {\xlor{\xl[1]}{\xl[2]} : \ltype{\ralt{\mathR[11]}{\mathR[21]}}{\ralt{\mathR[12]}{\mathR[22]}}}
\end{mathpar}
    \caption{Typing rules for core bijective lenses}
    \label{fig:core-ltype}
\end{figure}

\subsection{Summary}
While well-typed bijective lenses provide strong correctness guarantees, they have limited expressiveness. They can only translate between formats that are represented by regular expressions. Our contribution, match-reference bijective lenses, extends bijective lenses to make them more expressive, while maintaining the same strong correctness guarantees. We extend two aspects of the core bijective lenses: their constructors and their type system. Our extension of the type system includes an extension to the regular expressions that make up a lens type. So, we set the foundation for our discussion of match-reference bijective lenses by first presenting our extension to regular expressions: match-reference regular expressions.

\section{Match-Reference Regular Expressions}
\label{sec:mre}

\newcommand{\qstr}[1]{``#1''}

\subsection{Introducing match-reference regular expressions}

\begin{figure}
\begin{subfigure}{\textwidth}
\centering 
\includegraphics[width=\textwidth]{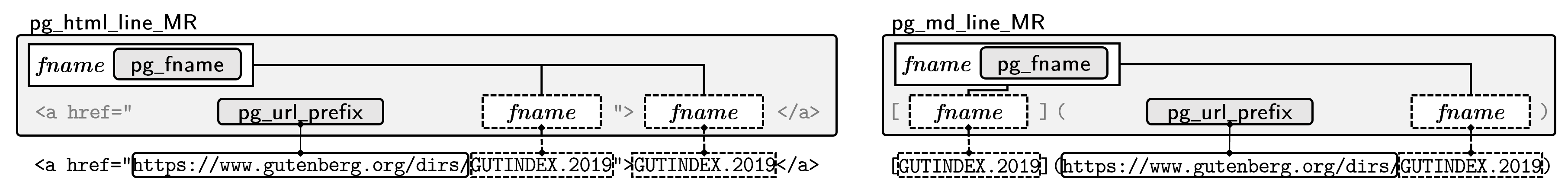}
\caption{The match-reference regexes \xrdef{pg\_html\_line\_MR} and \xrdef{pg\_md\_line\_MR} applied to a valid Project Gutenberg hyperlink in HTML and Markdown, respectively. In both cases, the variable, \xrvar{fname}, matches the same string, \xlit{GUTINDEX.2019}, which is in the language of \xrdef{pg\_fname}.}
\label{fig:mr_html_md_eg}
\end{subfigure}
\begin{subfigure}{\textwidth}
\centering
\includegraphics[width=.55\textwidth]{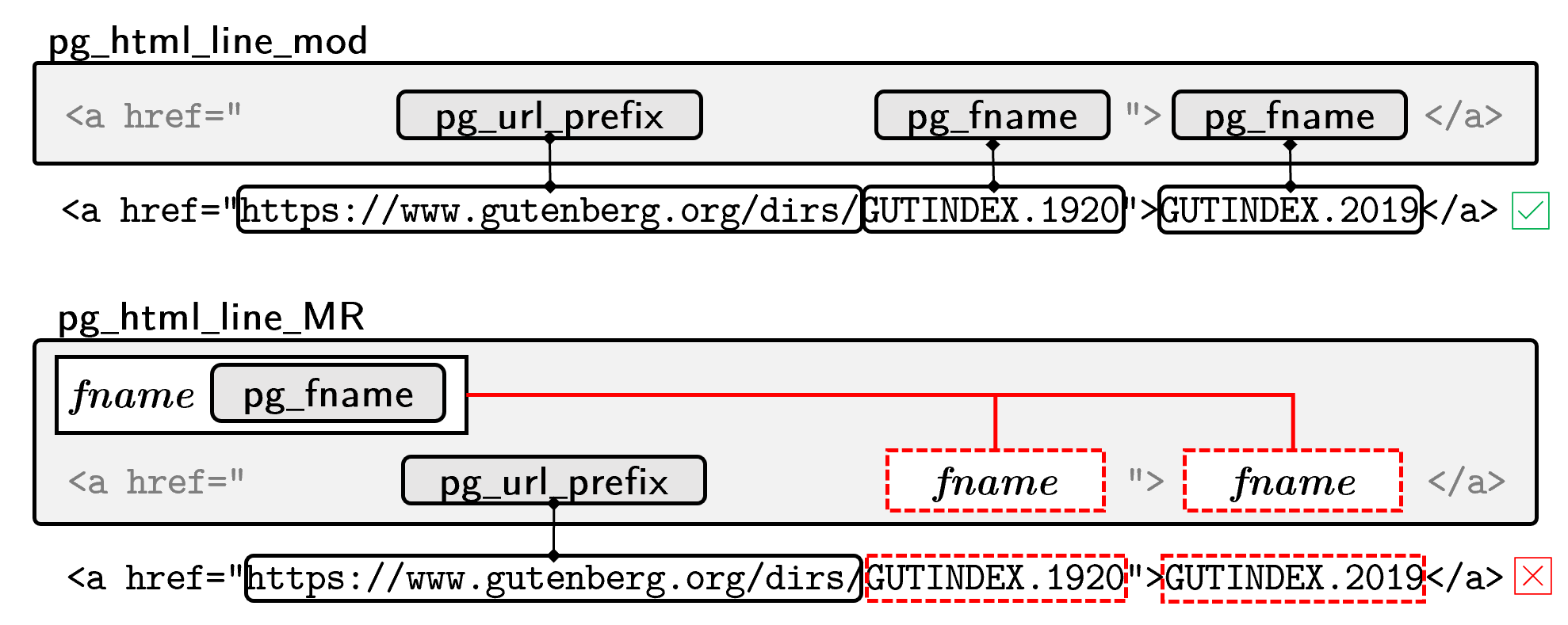}
\caption{Core regex \xrdef{pg\_html\_line\_mod} versus corresponding match-reference regex \xrdef{pg\_html\_line\_MR} applied to an invalid Project Gutenberg HTML hyperlink. The core regex accepts the badly formatted hyperlink, given that both \xlit{GUTINDEX.1920} and \xlit{GUTINDEX.2019} are in the language of \xrdef{pg\_fname}. The match-reference regex does not accept the hyperlink because, while they are both in the language of \xrdef{pg\_fname}, \xlit{GUTINDEX.1920} and \xlit{GUTINDEX.2019} don't match.}
\label{fig:mr-vs-core}
\end{subfigure}
\caption{Applying match-reference regexes to Project Gutenberg hyperlinks}

\end{figure}

Match-reference regular expressions allow us to represent formats with internal dependencies that cannot be represented by core regular expressions. For instance, the Project Gutenberg hyperlinks given in \autoref{fig:gut-ex-both} are structured such that the displayed text, \texttt{GUTINDEX.\#\#\#\#}, matches the name of the file pointed to by the hidden url. Any core regular expression that represents the hyperlinks, such as those given in \autoref{fig:pg-regex-summary}, will be an impoverished representation because they cannot represent this internal dependency. With match-reference regular expressions, on the other hand, we \emph{can} capture this structure and exclude badly formatted hyperlinks where the displayed text and url do not match. 

We will show how to use match-reference regular expressions by crafting one that represents the file formats in \autoref{fig:gut-ex-both}. First, given that the filename is the part that is repeated, we define two regexes, one that represents the filename's format, and one that represents just the prefix of the url:
\begin{align*}
\xrdef{pg\_fname} &:= \xlit{GUTINDEX.}[\xlit{0-9}][\xlit{0-9}][\xlit{0-9}][\xlit{0-9}] \\
\xrdef{pg\_url\_prefix} & := \xlit{https://www.gutenberg.org/dirs/}
\end{align*}
Next, we represent a single line of the HTML and Markdown formats. Here, we use the match-reference extensions to the core regular expressions: variables, \dvarx, and binding expressions, \bindin{\dvarx}{\mathR[1]}{\mathR[2]}. The binding expression gives the variable, \dvarx, a type, \mathR[1]. When \dvarx\ appears in \mathR[2], the body of the binding expression,  not only must \dvarx\ match a string in the language of \mathR[1], but \dvarx\ must match the same string wherever it appears. So, a line of the Project Gutenberg hyperlinks can be written as:
\begin{align*}
\xrdef{pg\_html\_line\_MR} &:=
    \bindin{\xrvar{fname}}
        {\xrdef{pg\_fname}}
            {\rand{\rand{\xlit{<href="}}{\rand{\xrdef{pg\_url\_prefix}}{\rand{\xrvar{fname}}{\xlit{">}}}}}
                  {\rand{\xrvar{fname}}{\xlit{</a>}}}}
                  \\
\xrdef{pg\_md\_line\_MR} &:=
    \bindin{\xrvar{fname}}
           {\xrdef{pg\_fname}}
           {\rand{\rand{\xlit{[}}{\rand{\xrvar{fname}}{\xlit{]}}}}{\rand{\xlit{(}}{\rand{\rand{\xrdef{pg\_url\_prefix}}{\xrvar{fname}}}{\xlit{)}}}}}
\end{align*}

In both \xrdef{pg\_html\_line\_MR} and \xrdef{pg\_md\_line\_MR}, the variable \xrvar{fname} is given the type \xrdef{pg\_fname} and then appears twice in the body of the binding expression, where it has to match the same string (illustrated in \autoref{fig:mr_html_md_eg}).

At this point, we can contrast a named regex, like \xrdef{pg\_fname}, and a match-reference variable, like \xrvar{fname}. Say we refactor \xrdef{pg\_html\_line} so that it uses \xrdef{pg\_fname} wherever the filename appears:

\begin{center}
\xrdef{pg\_html\_line\_mod} :=
 \rand{\rand{\xlit{<a href="}}{\rand{\rand{\xrdef{pg\_url\_prefix}}{\xrdef{pg\_fname}}}{\xlit{">}}}}
                                     {\rand{\xrdef{pg\_fname}}{\xlit{</a>}}}
\end{center}
Wherever a string matches \xrdef{pg\_fname} in \xrdef{pg\_html\_line\_mod}, it \emph{only} has to be in the language of \xrdef{pg\_fname}. So, the bad Project Gutenberg hyperlink
\begin{center}
\xlit{<href="https://www.gutenberg.org/dirs/GUTINDEX.1920">GUTINDEX.2019</a>}
\end{center}
is in the language of \xrdef{pg\_html\_line\_mod}, but it is \emph{not} in the language of \xrdef{pg\_html\_line\_MR} as illustrated in \autoref{fig:mr-vs-core}. That is, as a match-reference expression, \xrdef{pg\_html\_line\_MR} can enforce what a simple regular expression cannot: the displayed filename and the linked file should match.

Finally, we represent the list of hyperlinks by starring the individual line:
\begin{align*}
\xrdef{pg\_html\_MR} & := \rstar{\xrdef{pg\_html\_line\_MR}} \\
\xrdef{pg\_md\_MR} & := \rstar{\xrdef{pg\_md\_line\_MR}}
\end{align*}
Given that in both \xrdef{pg\_html\_MR} and \xrdef{pg\_md\_MR} the binding form  \bindin{\xrvar{fname}}{\mathR[1]}{\mathR[2]} appears under the star, \xrvar{fname} can match a different filename on each line. Within a line, however, \xrvar{fname} must still match the same name. 

\subsection{Syntax and Denotational Semantics}

The syntax for match-reference regexes is given in \autoref{fig:mre-syntax}. As mentioned earlier, regular expressions are augmented with variables, \dvarx\ \xin \emath{V},  where \emath{V} is a set of variables that is disjoint from the input alphabet \emath{\Sigma}. The binding form, \bindin{\dvarx}{\mathR[1]}{\mathR[2]}, sets the \mathR[1] as the type of the variable \dvarx.

A \ename\ regular expression, \mathR, denotes a set of strings.
The semantics of \mathR, \rL{\mathR}, is given in \autoref{fig:mre-sem},  where \renv\ maps variables to strings. The semantics is largely standard for regular expressions, aside for the two new forms: variables and variable binding.
The language of a variable \dvarx\ is a single string, \set{\s}, when \dvarx\ is in the domain of \renv, assuming that \s\ = \lu{\renv}{\dvar} and that looking up \dvar\ in a given environment \renv\ is deterministic.
The language of the binding form \bindin{\dvar}{\mathR[1]}{\mathR[2]} is the set of strings denoted by \mathR[2] in the set of environments that extend \renv\ with a mapping of \dvar\ to \xpr{\s}, where \xpr{\s} is any string in \rL{\mathR[1]}.

\begin{example}[\mathR\ = \bindin{\dvarx}{\rstar{\rconst{a}}}{\rand{\dvarx}{\rand{\rconst{b}}{\dvarx}}}]
\label{ex:var-bind-semantics}

\mathR\ in the empty environment \empenv\ denotes: \set{\rconst{b}, \rconst{aba}, ..., \rconst{a}^i\rconst{b}\rconst{a}^i, ...}. The string \qstr{aba} is in \rL[\empenv]{\mathR} because \qstr{a} \xin \rL[\empenv]{\rstar{\rconst{a}}} and \qstr{aba} \xin \rL[\set{\xpair{\dvarx}{\rconst{a}}}]{\rand{\dvarx}{\rand{\rconst{b}}{\dvarx}}}. More generally: \s\ \xin \rL[\empenv]{\bindin{\dvarx}{\rstar{\rconst{a}}}{\rand{\dvarx}{\rand{\rconst{b}}{\dvarx}}}} if and only if \xpr{\s} \xin \rL[\empenv]{\rstar{\rconst{a}}} and \s\ \xin \rL[\set{\xpair{\dvarx}{\s'}}]{\rand{\dvarx}{\rand{\rconst{b}}{\dvarx}}}.

\end{example}

It follows from the semantics that the binding form, \bindin{\dvar}{\mathR[1]}{\mathR[2]} not only sets \mathR[1] as the type of the variable, \dvarx, it also limits the scope of \dvarx\ to \mathR[2]. We can see the effects of scoping when a binding expression, such as the expression in \autoref{ex:var-bind-semantics}, is starred.

\begin{example}[\mathR\ = \rstar{\bindin{\dvarx}{\rstar{\rconst{a}}}{\rand{\dvarx}{\rand{\rconst{b}}{\dvarx}}}}]
 \mathR\ in the empty environment \empenv, denotes: \set{\lambda, \rconst{b}, \rconst{aba}, \rconst{bb}, \rconst{baba}, \rconst{abab}, \\ \rconst{abaaba}, \rconst{aabaa}, ..., (\rconst{a}^{i_1}\rconst{ba}^{i_1}) \stackrel{k-2}{\cdots} (\rconst{a}^{i_k}\rconst{ba}^{i_k}), ...}.
Broadly speaking, the expression under a star is iterated \emath{k} times, so, in this case, \bindin{\dvarx}{\rstar{\rconst{a}}}{\rand{\dvarx}{\rand{\rconst{b}}{\dvarx}}} is used \emath{k} times to match, or generate, a sub-string. For instance, \qstr{abab} \xin \rL[\empenv]{\mathR} because \qstr{aba} \xin \rL[\set{\xpair{\dvarx}{\rconst{a}}}]{\rand{\dvarx}{\rand{\rconst{b}}{\dvarx}}} and \qstr{b} \xin \rL[\set{\xpair{\dvarx}{\lambda}}]{\rand{\dvarx}{\rand{\rconst{b}}{\dvarx}}} Within a single iteration, \dvar\ matches the same string, \xpr{\s},  where \xpr{\s} \xin \rL[\empenv]{\rstar{\rconst{a}}}. Across iterations, however, the match for \dvarx\ is reset.
\end{example}
\begin{figure}[t]
    \begin{subfigure}{\textwidth}

    \input{definitions/MRE_grammar}

    \caption{\upename\ Regex Syntax} \label{fig:mre-syntax}

    \end{subfigure}
    \begin{subfigure}{\textwidth}
    \input{definitions/MRE_Semantics_Simple}
    \caption{\upename\ Regex Semantics} \label{fig:mre-sem}
    \end{subfigure}
    \caption{Formalisms for match-reference regexes}
\end{figure}

\subsection{Contrasting match-reference regexes and back-reference regexes }
\label{ssec:mre-vs-bref}

\subsubsection{Characterizing back-references}
Most scripting languages make use of regular expressions and often include features that go beyond the core regular expressions. Many of these features are found in the popular PCRE (perl-compatible regular expression) library. Specifically, match-reference regular expressions are analogous to the extension known as \emph{back-references} \cite{PCREman}. Back-references augment regular expressions with the ability to specify languages that have repeating sub-strings. A predominant version of the back-reference syntax indexes a parenthesized sub-expression based on where it occurs in the expression, and then refers to that match with its index: \emath{\backslash n}. So, \emath{n} is set when then \emath{n^{th}} parenthesized expression is matched, and then the match to \emath{n} is reused wherever \emath{\backslash n} appears. Alternative forms explicitly gives names to sub-expressions, and use those names to indicate a repeated match elsewhere in the expression. For the purposes of this paper, we adopt the following generic syntax that uses names:

\begin{enumerate}
    \item \emath{\{x=r \}}: variable definition, matches a string, \s\ in the language of \emath{r}, and sets the variable \dvarx\ to \s\
    \item \emath{x}: variable reuse, matches the string, \s\, that \dvarx\ was set to. 
\end{enumerate}

\begin{example}[\emath{r} = \rand{\{ \dvarx = \rstar{\tta} \}}{\rand{\ttb}{\dvarx}}]
\emath{r} represents the language \set{\ttb, \tta\ttb\tta, ..., \tta^k\ttb\tta^k, ...}. When \emath{r} is used to match a string, \emath{\{ \dvarx = \rstar{\tta} \}} matches string prefixes that are \emath{\tta^k} and sets \dvarx\ to \emath{\tta^k}. Then, \ttb\ matches \ttb, as expected. The last part of r, \dvarx, matches the string \dvarx\ was set to earlier: \emath{\tta^k}. Note that, in the pre-dominant syntax this expression could be written as \rand{(\rstar{\tta})}{\rand{\ttb}{\backslash 1}}.
\end{example}

\subsubsection{Back-references: tricky semantics} 

When back-references are added to regular expressions, some instances that include Kleene stars or alternations lack straight-forward interpretations \cite{Campeanu2003, Schmid2012}. There are two sources of divergent interpretations: a variable referring to an unmatched expression or a variable referring to an expression with several matches.

In the first case, when a variable refers to an unmatched expression, it gives rise various possible interpretations: for instance, the variable refers to the empty string or to the empty set. Depending on the context, one of these interpretations may be preferable to the other, but the predominant approach is the first \cite{Schmid2012}.
\begin{example}[\emath{r} = \emath{\rand{(\ralt{\set{x = \tta^*}}{\set{y = \ttb^*}})}{\rand{\rand{\ttc}{\dvarx}}}{\rand{\ttc}{y}}}]
\label{ex:backref-no-matches}
Under one interpretation the language of \emath{r} is \set{\ttc\ttc, \tta\ttc\tta\ttc, \ttb\ttc\ttc\ttb, \cdots, \tta^k\ttc\tta^k\ttc, \ttb^k\ttc\ttc\ttb^k, \cdots}. Under the other interpretation, the language of \emath{r} is the empty set, because either \dvarx\ or \emath{y} will always be unmatched
\end{example}

In the second case, when a variable refers to an expression with several matches, implementations generally choose to use either the first match or the last match.
\begin{example}[\emath{r} = \rand{\rstar{(\rand{\set{x = \rstar{\tta}}}{\rand{\ttb}{\dvarx}})}}{\rand{\ttc}{\dvarx}}]
\label{ex:backref-many-matches}
In \emath{r},  \dvarx\ is defined within a starred expression and the variable \dvarx\ appears both within and outside of that starred expression. The variable under the star lends itself to the interpretation that its match gets reset each time the starred expression gets evaluated. The variable outside the star though, occurs after the match for \dvarx\ has been reset \emath{n} times under the star, which gives rise to two predominant interpretations: either the external \dvarx\ matches the first or last string that was matched by the internal \dvarx. Under the first interpretation the language of \emath{r} is: \defset{\tta^{n_1}\ttb\tta^{n_1} \cdot \cdots \cdot \tta^{n_m}\ttb\tta^{n_m}\cdot \ttc \cdot \underline{\tta^{n_1}}}{0 < m, 0 < n_k}  (\xunion \set{\ttc} depending on how we deal with unmatched expressions). Under the second interpretation, the language of \emath{r} is: \defset{\tta^{n_1}\ttb\tta^{n_1} \cdot \cdots \cdot \tta^{n_m}\ttb\tta^{n_m}\cdot \ttc \cdot \underline{\tta^{n_m}}}{0 < m, 0 < n_k} (\xunion \set{\ttc} depending on how we deal with unmatched expressions)
\end{example}

So, the syntax for back-references, in its many flavors, has several reasonable but divergent 
interpretations, giving rise to a proliferation of semantics. By teasing apart the different components of the back-references, and making the syntax more explicitly correspond to these separate components, match-references regular expressions have a clearer semantics than back-references regular expressions


\subsubsection{Variables in back-reference regexes and match-reference regexes}

The semantics of a variable in back-reference regexes depends on three components:
\begin{enumerate}
    \item the variable's type: what is the language of the strings that match the variable
    \item the variable's location: which sub-strings match the variable
    \item the variable's scope: where in the expression can the variable meaningfully appear
\end{enumerate}

The mainstream formalism for back-references syntactically marks two of these aspects: giving a type to a variable and using variables to indicate the location of string repetitions. However, back-references in their many forms typically include an expression that does double duty: both setting the type of a variable, and matching an element of the string to give the variable a value. This double-duty expression, arguably, makes it more difficult to edit back-references: swapping sub-expressions might put a variable before its definition, for instance. In contrast, match-references keep these two roles syntactically distinct. They have a syntactic form that is solely dedicated to indicating the type of a variable (in keeping with how we might define function parameters or declare variable types, for instance). A second form, the variable, is exclusively used to mark the parts of an expression where a string should match that variable. By keeping the variable definition and use separate, match-reference regexes lend themselves to writing expressions that are clearer and easier to edit than back-reference expressions.

The third aspect, the variable's scope, is generally not explicitly indicated in the syntax of back-references. The defaults seem to be that either a variable is in scope for the entire expression or it is in scope after its definition form appears in the expression. This lack of explicit scope is part of what leads to the ambiguity in \autoref{ex:backref-many-matches}: \rand{\rstar{(\rand{\set{x = \rstar{\tta}}}{\rand{\ttb}{\dvarx}})}}{\rand{\ttc}{\dvarx}}. The third appearance of \dvarx\ is outside of the star under which it is given a type, and therefore, it is ambiguous which string that third \dvarx\ should match. A clearer version of this back-reference extracts the instance which the third \dvarx\ should match, with an additional expression:  \rand{\rstar{(\rand{\set{\dvarx[1] = \rstar{\tta}}}{\rand{\ttb}{\dvarx[1]}})}}{(\rand{\rand{\set{\dvarx[2] = \rstar{\tta}}}{\rand{\ttb}{\dvarx[2]}}}{\rand{\ttc}{\dvarx[2]}})}.

The binding form of the match-reference regex, \bindin{\dvarx}{\mathR[1]}{\mathR[2]} readily lends itself to the interpretation of setting the scope of the variable within the body of \mathR[2]. This means that everywhere \dvarx\ appears in \mathR[2], it will match the same string. It is this property, the explicit marking of scope, that allows us to keep the expressivity of back-references, while extracting the definition of a variable from its use.

The expressivity is maintained because the location of the definition section of the binding form, \emath{\dvarx : \mathR[1]}, is meaningful. Moving its location affects what a variable will match. This is most apparent when the binding form appears under the star: \rstar{\bindin{\dvarx}{\mathR[1]}{\mathR[2]}}. In this expression, \dvarx\ goes out of scope at the end of each iteration. Once it enters scope again, \dvarx\ may match a different string. If we move the definition outside of the star, \bindin{\dvarx}{\mathR[1]}{(\rstar{\mathR[2]})}, \dvarx\ does not go out of scope after each iteration, so each occurance of \dvarx, across iterations of \mathR[2], will match the same string.

In addition, the match-reference regex syntax makes it harder to write ambiguous expressions like \rand{\rstar{(\rand{\set{x = \rstar{\tta}}}{\rand{\ttb}{\dvarx}})}}{\rand{\ttc}{\dvarx}}. The naive translation:  \rand{\rstar{\bindin{\dvarx}{\rstar{\tta}}{\rand{\rand{\dvarx}{\ttb}}{\dvarx}}}}{\rand{\ttc}{\dvarx}} actually denotes the empty language. In the naive translation, the final occurance of \dvarx\ occurs outside the scope of any definition for \dvarx\ and this is why it denotes the empty language. When using match-reference regexes, for better or worse, we must translate the back-reference expression into an unambiguous form, such as \rand{\rstar{\bindin{\dvarx}{\rstar{\tta}}{\rand{\rand{\dvarx}{\ttb}}{\dvarx}}}}{\bindin{\dvarx}{\rstar{\tta}}{\rand{\rand{\rand{\dvarx}{\ttb}}{\dvarx}}{\rand{\ttc}{\dvarx}}}}. The explicit marking of scope in match-reference regexes may lead us to write longer expressions, but the resulting expressions don't have some of the ambiguities inherent in back-references.

Match-reference regexes, and their automata (discussed below in \autoref{sec:mras}), are inspired by past work on formalizing back-reference regexes which originated with Campeanu et al. \cite{Campeanu2009a, Campeanu2009b, Yu2004} and continues in work by Schmid and Freydenberger \cite{Freydenberger2018, Schmid2012,Schmid2019}. Campeanu et al.'s initial work gave rise to pattern expressions and pattern automata. Pattern expressions, like match-reference expressions, separate the definition of a variable's type from the use of that variable. However, pattern expressions are not as expressive as back-references, because the binding form is top-level and therefore cannot be put under the star. The subsequent works provide well-defined semantics and automata for formalisms that more closely mirror the structure of back-references, insofar as these formalisms have a syntactic form that does the double duty of giving a variable its type and finding a match for that variable. Our contribution, match-reference regexes and their automata, are like pattern expressions, because they tease apart the form that gives variables their types and the form that finds a match for variables but, like the subsequent work we draw on, their expressiveness is akin to that of back-references. 
 
\section{Match-Reference Bijective Lenses}
\label{sec:mre-lens}

\subsection{Introducing match-reference bijective lenses}

\begin{figure}
\centering 

\includegraphics[width=\textwidth]{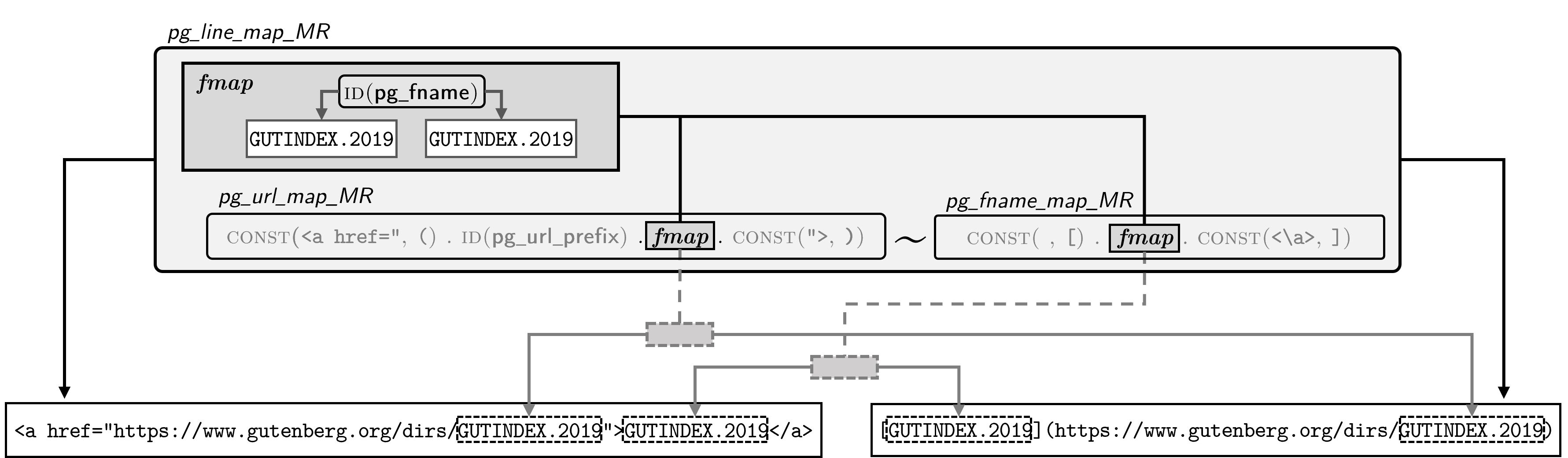}
\caption{Applying \xldef{pg\_line\_map\_MR} to translate a Project Gutenberg hyperlink between HTML and Markdown. The variable lens, \xrvar{fmap}, translates the same strings wherever it appears in the body of \xldef{pg\_line\_map\_MR}'s link lens. The translation \xrvar{fmap} performs is defined by the lens: \xid{\xrdef{pg\_fname}}.}
\label{fig:mr_lens_eg}
\end{figure}
Bijective \ename\ lenses are designed to translate between formats with the internal dependencies that can be represented by \ename\ regexes. In contrast, the core bijective lenses can only translate between formats represented by regular expressions. For instance, we cannot write a simple bijective lens between the two formats in \autoref{fig:gut-ex-both} that will reject badly formatted hyperlinks where the visible text doesn't correspond to the underlying filename. We \emph{can} write a match-reference bijective lens that will only translate well-formatted Project Gutenberg hyperlinks, and that will ensure that the underlying url and displayed text remain in sync during the transformation. 

Match-reference lenses extend the core bijective lenses with a variable expression, \xlvar, and a link expression, \xlab{\xlvar}{\xl[1]}{\xl[2]}. In the link expression, the variable is typed by the lens \xl[1] and used in \xl[2]. Wherever \xlvar\ appears in the body, \xl[2], it translates the same input string into the same output string, where that pair of input and output strings must be a translation performed by \xl[1]. 

So, using the variable and link constructors, we can write a match-reference lens that  translates a single line of hyperlinks between HTML and Markdown.
First, we write two lenses that will translate between the url and visible text, respectively. 
\begin{align*}
    \xldef{pg\_url\_map\_MR} & := \xconst{\xlit{<href="}}{\xlit{(}} \ . \ \xid{\xrdef{pg\_url\_prefix}} \ . \ \xrvar{fmap} \ . \ \xconst{\xlit{">}}{\xlit{)}} \\
    \xldef{pg\_fname\_map\_MR} & := \xconst{ \ }{ \xlit{[}} \ . \  \xrvar{fmap} \ . \ \xconst{\xlit{</a>}}{\xlit{]}}
\end{align*}
In both lenses, we use the variable lens, \xrvar{fmap}, to transform the filename but at this point \xrvar{fmap} is not well-defined because it has not been bound in the body of a link lens. So, next, we write a link lens that binds \xrvar{fmap} and translates a single hyperlink:

\begin{center}
\xldef{pg\_line\_map\_MR} := \xlab{\xrvar{fmap}}{\xid{\xrdef{pg\_fname}}}{\sxswap{\xldef{pg\_url\_map\_MR}}{ \xldef{pg\_fname\_map\_MR}}}
\end{center}

\noindent Everywhere the variable lens \xrvar{fmap} appears in the body of \xldef{pg\_line\_map\_MR}'s  link lens, not only does it apply the lens \xid{\xrdef{pg\_fname}}, it will translate the same input string into the same output string, as illustrated in \autoref{fig:mr_lens_eg}. So this lens can only be applied to hyperlinks where the displayed text matches the name of the underlying file, and it will only produce entries where this correpondence between the visible and hidden part of the hyperlink holds.

Finally, using the iteration lens from the core set of bijective lenses, we build a lens that translates a list of hyperlinks between HTML and Markdown.
\begin{center}
\xldef{pg\_map\_MR} := \xiter{\xldef{pg\_line\_map\_MR}}
\end{center}
Given that the link lens \xldef{pg\_line\_map\_MR} is iterated here, the translation performed by \xrvar{fmap} gets reset after each iteration. So, within a hyperlink, \xrvar{fmap} transforms the same input string to the same output string, where that transformation is the result of applying \xid{\xrdef{pg\_fname}} to the input string, but across entries the input and output strings translated by \xrvar{fmap} can differ.

\subsection{Syntax and Denotational Semantics}
\begin{figure}

\begin{subfigure}{\textwidth}
\input{definitions/MRLens_Grammar}
\caption{MR lens grammar}
\label{fig:MRL-Grammar}
\end{subfigure}
\begin{subfigure}{\textwidth}

\input{definitions/MRLens_DenotationalSemantics}
\caption{MR lens denotational semantics}
\label{fig:MRL-DenSem}
\end{subfigure}
\caption{Bijective match-reference lens (MR lens) formalisms}
\end{figure}
The grammar for bijective match-reference lenses is given in \autoref{fig:MRL-Grammar}.
To re-iterate, we extend the core bijective lenses with two new forms: the variable lens, \xlvar\, and the link lens constructor, \xlab{\xlvar}{\xl[1]}{\xl[2]}. When describing a link lens, we say that the lens \xl[1] is linked to \xlvar\ in \xl[2]. Notice that this notation mirrors how we extend regular expressions in match reference regular expressions. The \xid{\mathR} form has a subtler change: its parameter is a match-reference regex, rather than a basic regex.

The denotational semantics for match-reference lenses are given in \autoref{fig:MRL-DenSem}. The denotation of a lens \xl, is a set of string pairs. If \xpair{\s[1]}{\s[2]} is in the denotation of \xl\ then we say that \xl\ \emph{maps between} \s[1] and \s[2]. We evaluate the denotation of the match-reference lenses in the context of an environment, \lenv, that maps variables to string pairs.

The semantics of the core bijective lenses are largely unchanged given that they do not reference or update the lens value environment, \lenv. The semantics of the id lens, \xid{\mathR}, changes because it is paramatrized by a match-reference regex, rather than a basic regular expression. As such, the denotation of \xid{\mathR} uses the semantics for match-reference regexes given in \autoref{fig:mre-sem}, which uses the regex value environment, \renv, that maps variables to strings. Note that the semantics of \xid{\mathR} states that the denotation of \mathR\ is evaluated in the empty environment. This means that the denotation of \mathR\ does not refer to any values outside of those that occur while evaluating \mathR\ itself.

The denotation of a variable lens can only be specified relative to \lenv, and \lenv\ is only updated by the link lens. So, the link and variable lenses work together, via \lenv, to ensure that equal substrings will be translated the same way, so that the corresponding substrings remain equal after the format has been transformed.

\begin{example}[\xl\ = \xlab{\xlvar}{\xiter{\xconst{\rconst{a}}{\rconst{A}}}}{\xlvar \ . \ \xid{\rconst{b}} \ . \ \xlvar}]
\xl\ in the empty environment, \empenv, denotes \set{\xpair{\rconst{b}}{\rconst{b}}, \xpair{\rconst{aba}}{\rconst{AbA}}, ...,  \xpair{\rconst{a}^{k}\rconst{b}\rconst{a}^k}{\rconst{A}^k\rconst{b}\rconst{A}^k}, ...}. So, a string pair \xpair{\rconst{a}^{i}\rconst{b}\rconst{a}^j}{\rconst{A}^i\rconst{b}\rconst{A}^j} is in the language of \xl\ only if \emath{i = j}. In particular, the string pair \xpair{\rconst{aabaa}}{\rconst{AAbAA}} is in \denotes[\empenv]{\xl} because \xpair{\rconst{aa}}{\rconst{AA}} is in \denotes[\empenv]{\xiter{\xconst{\rconst{a}}{\rconst{A}}}}  and \xpair{\rconst{aa}}{\rconst{AA}} is in \denotes[\set{\xpair{\xlvar}{\xpair{\rconst{aa}}{\rconst{AA}}}}]{y} which means that \xpair{\rconst{aabaa}}{\rconst{AAbAA}} is in \denotes[\set{\xpair{\xlvar}{\xpair{\rconst{aa}}{\rconst{AA}}}}]{\xlvar \ . \ \xid{\rconst{b}} \ . \ \xlvar}.
\end{example}

We can think of the string pairs in the denotation of a lens as the result of applying a lens' \lget\ and \lput\ function. So, if \xpair{\s[1]}{\s[2]} \xin \denotes{\xl} then \xl.\lget(\s[1]) = \s[2] and \xl.\lget(\s[2]) = \s[1] in the context of \lenv.

\begin{example}[\xl\ = \xlab{\xlvar}{\xiter{\xconst{\rconst{a}}{\rconst{A}}}}{\xlvar \ . \ \xid{\rconst{b}} \ . \ \xlvar}] In the empty environment, \empenv, \xl.\lget(\emath{\rconst{a}^{k}\rconst{b}\rconst{a}^k}) = \emath{\rconst{A}^k\rconst{b}\rconst{A}^k} and \xl.\lput(\emath{\rconst{A}^k\rconst{b}\rconst{A}^k}) = \emath{\rconst{a}^{k}\rconst{b}\rconst{a}^k}.  Say we apply \xl's \lget\ function to the string \qstr{\rconst{aabaa}} in \empenv. We first need to translate the prefix \qstr{\rconst{aa}} with \xlvar's \lget\ function. However, no translation exists for \xlvar\ in the empty lens value environment, \empenv. So, we use \xiter{\xconst{\rconst{a}}{\rconst{A}}}'s \lget\ function to translate \qstr{\rconst{aa}} to \qstr{\rconst{AA}}, and store that translation, \xpair{\rconst{aa}}{\rconst{AA}}, in \lenv. Next, \qstr{\rconst{b}} is translated to \qstr{\rconst{b}} with the identity lens' \lget\ function, given that \qstr{\rconst{b}} is in the language of the match-reference regex \rconst{b} evaluated in the empty environment. Finally, we translate the suffix \qstr{\rconst{aa}} with \xlvar's \lget\ function in the environment \set{\xpair{\xlvar}{\xpair{\rconst{aa}}{\rconst{AA}}}}. This time a translation exists for \xlvar\ in the environment, so it translates \qstr{\rconst{aa}} into \qstr{\rconst{AA}} a second time.
\end{example}

\newcommand{\tyrule}[1]{\textsc{#1}}
\subsection{Type System}

Just like core bijective lenses, bijective match-reference lenses have a type system within which a well-typed lens is guaranteed to only perform good translations. The typing judgment is \ljudg{\xl}{\ltype{\mathR[1]}{\mathR[2]}} where \ltenv\ and \rtenv\ are type environments, \xl\ is a match-reference lens, and \mathR[1] and \mathR[2] are match-reference regular expressions. The \ltenv\ type environment maps variables to lens types and the \rtenv\ type environment maps variables to regular expressions. The typing rules include two side-conditions: one puts constraints on the type environments and the other puts constraints on the match-reference regexes that appear in lens types. We give the side conditions on environments in \autoref{sssec:wf-envs} and we discuss the constraints on match-reference regexes in \autoref{sssec:unambiguity}. With those definitions in place, we present the typing rules in \autoref{sssec:tyrules}

\subsubsection{Well-formed type environments}
\label{sssec:wf-envs}

To re-iterate: the \ltenv\ type environment maps variables to lens types and the \rtenv\ type environment maps variables to regular expressions. More precisely, the type of a lens variable, \xlvar, is always a pair of match-reference regex variables, \ltype{\dvarx[1]}{\dvarx[2]}. The regex type environment, \rtenv, in turn maps these variables, \dvarx[1] and \dvarx[2], to \emph{their} types, \mathR[1] and \mathR[2]. 

Broadly speaking, \rtenv\ is well-formed if the free variables of all the regexes bound in \rtenv\ are in the domain of \rtenv. In other words, if \mathR\ is bound \rtenv and \rtenv is well-formed, then the free variables in \mathR\ are a subset of \rtenv's domain. A lens type environment, \ltenv, can only be well-formed in conjunction with well-formed regex type environment, \rtenv, and only  if all of the regex variables in the types defined in \ltenv\ appear in \rtenv's domain. Below, we formally define the three properties that work together to encompass what it means for the type environments to be well-formed.

First, we define what it means for a regex type environment to be good for a regex:

\begin{definition}[\emath{\rtenv \vdash \mathR}: \rtenv\ is good for \mathR]
\label{def:rt-gf-R}
\leavevmode

\noindent
A regex type environment, \rtenv, is good for a match-reference regex, \mathR, if all the free variables in \mathR\ exist in the domain of \rtenv. Formally:
\begin{mathpar}
    \inferrule{\xfa \dvar \xin \fvs{\mathR}, \dvar \xin \xdom{\rtenv}}
                {\rtenv \vdash \mathR}
\end{mathpar}
\end{definition}

Next, we use \autoref{def:rt-gf-R} to define what it means for a regex type environment to be well-formed:

\begin{definition}[\emath{\rtenv \vdash *}: \rtenv\ is well-formed]
\label{def:good-rtenv}
\leavevmode

\noindent
The empty regex type environment is well-formed. The result of binding the \dvarx\ to \mathR, in a well-formed environment, \rtenv, is also well-formed, as long as \dvarx\ is a new variable, and \rtenv\ is good for \mathR, that is, all the free variables in \mathR\ are already in the domain of \rtenv. Formally:
    \begin{mathpar}
        \inferrule{ }{ \empenv \vdash *}

        \inferrule{ \rtenv \vdash * \and \dvar \xnin \xdom{\rtenv} \and \rtenv \vdash \mathR}
                    {\extenv{\rtenv}{\dvar}{\mathR} \vdash *}
    \end{mathpar}
\end{definition}

Finally, we can define what it means for the two type environments, together, to be well-formed:

\begin{definition}[\emath{\ltenv, \rtenv \vdash *}: \ltenv\ and \rtenv\ are  well-formed]
\label{def:good-lrtenvs}
\leavevmode

\noindent
If the lens and regex type environments, \ltenv\ and \rtenv\ respectively, are well formed then the bound variables in each environment are unique. Moreover, for each lens variable, \xlvar, in the domain of \ltenv, if \lu{\ltenv}{\xlvar} = \ltype{\dvarx[1]}{\dvarx[2]}, then \dvarx[1], \dvarx[2] \xin \xdom{\rtenv}. Finally, because \rtenv\ is well-formed, for each regex, \mathR, in the range of \rtenv, a prefix of \rtenv\ is good for \mathR. In other words, for any given variable, \dvarx, in the domain of \rtenv, we know that \rtenv\ = \xpr{\rtenv}, \xpair{\dvarx}{\mathR}, \xpr{\xpr{\rtenv}}, and that \xpr{\rtenv} \emath{\vdash} \mathR, and that, specifically, this property holds for the regexes bound to \dvarx[1] and \dvarx[2]. Formally:
    \begin{mathpar}
        \inferrule{ }{ \empenv, \empenv \vdash *}

        \inferrule{ \ltenv, \rtenv \vdash *
                        \and \xlvar \xnin \xdom{\ltenv}
                        \and \exte{\rtenv}{\mape{\dvarx[1]}{\mathR[1]}, \mape{\dvarx[2]}{\mathR[2]}} \vdash *}
                    {\exte{\ltenv}{\mape{\xlvar}{\ltype{\dvarx[1]}{\dvarx[2]}}}\exte{\rtenv}{\mape{\dvarx[1]}{\mathR[1]}, \mape{\dvarx[2]}{\mathR[2]}} \vdash *}
    \end{mathpar}
\end{definition}

\subsubsection{Strongly Unambiguous \upename\ Regexes}
\label{sssec:unambiguity}
   
A strongly unambiguous regex, \mathR, is an expression that parses each string in its language uniquely.
 
\begin{example}[{\mathR[1]} = \ralt{\rconst{ba}}{\rconst{ca}} vs {\mathR[2]} = \ralt{\rconst{a}}{\rconst{a}}]
\mathR[1] is an unambiguous regex, the two strings in its language are parsed uniquely by \mathR[1], for instance, \qstr{\rconst{ba}} only matches the left-hand branch of \mathR[1]. In contrast, \mathR[2] is not unambiguous, the string \qstr{\rconst{a}} can match either the left-hand or right-hand branch of \mathR[2].
\end{example}

As given in \autoref{fig:mre-sem}, the language of a match-reference regex is defined in the context of a regex value environment, \renv. However, we define unambiguity relative to a regex \emph{type} environment, \rtenv. To relate the two, we define the denotation of \rtenv\ as the set of value environments, \set{\renv[1], \renv[2], ...}, that are consistent with it, with the following two definitions: 

\begin{definition}[\emath{\rtenv \vdash \renv} : \renv\ is consistent with \rtenv]
\leavevmode

\noindent A regex value environment, \renv, is consistent with a regex type environment, \rtenv, if for any binding, \xpair{\dvarx}{\s}, in \renv, the string \s\ is in the language of \rL[\xpr{\renv}]{\mathR}, where:
    \begin{itemize}
        \item \renv\ = \xpr{\renv},(\dvarx, \s), \xpr{\xpr{\renv}},
        \item \rtenv\ = \xpr{\rtenv}, \xpair{\dvarx}{\mathR}, \xpr{\xpr{\rtenv}}, and
        \item \xpr{\renv} is consistent with \xpr{\rtenv}.
    \end{itemize}
\medspace

\noindent More formally:
    \begin{mathpar}
     			\inferrule{ }{\empenv \vdash \empenv}

     			\inferrule{\rtenv \vdash \renv \and \dvar \xnin \xdom{\rtenv}, \xdom{\renv} \and \s \xin \rL{\mathR}}
				  {\extenv{\rtenv}{\dvar}{\mathR} \vdash \extenv{\renv}{\dvar}{\s}}
     		\end{mathpar}
\end{definition}

\begin{definition}[\envD: Denotation of \rtenv]
\leavevmode

\noindent The denotation \rtenv\ is the set of value environments that are consistent with \rtenv. Formally: \envD\ = \defset{\renv}{\rtenv \vdash \renv}
\end{definition}

Next, as with the core regular expressions, unambiguous match-reference regexes are those that can be constructed from unambiguous versions of their operators. For core regular expressions, unambiguous operations are annotated with an exclamation point: \emath{op^{!}}. For match-reference expressions, unambiguous operations have an additional parameter: a regex type environment, \rtenv. We use the notation \emath{op^{!_{\rtenv}}} to say that \emath{op} is unambiguous in \rtenv. The unambiguous operations are defined as follows:

\begin{definition}[\emath{op^{!_{\rtenv}}}: Unambiguous operations]
\leavevmode

\noindent The outer-most operation of an expression is unambiguous in \rtenv\ if for any value environment, \renv, consistent with \rtenv, the parse for that operation is unique. More formally, we define each unambiguous operation as follows:
\begin{itemize}
    \item \emph{Unambiguous iteration}:
        \newline \uiter{\mathR} if \xfa \renv\ \xin \envD, \s[11], ..., \s[1n], \s[21] ,..., \s[2m] \xin \rL{\mathR}.
        \newline \mbox{~~~} if \s[11] \xdot ... \xdot \s[1n] = \s[21] \xdot ... \xdot \s[2m]
        \newline \mbox{~~~} then \emath{n = m} and \xfa \emath{k} from \emath{1} to \emath{n}, \s[1k] = \s[2k]

    \item \emph{Unambiguous concatenation}:
        \newline \uconcat{\mathR[1]}{\mathR[2]} if \xfa \renv\ \xin \envD, \s[11], \s[21] \xin \rL{\mathR[1]}, \s[12], \s[22] \xin \rL{\mathR[2]}.
        \newline \mbox{~~~} if \rand{\s[11]}{\s[12]} = \rand{\s[21]}{\s[22]}
        \newline \mbox{~~~} then \s[11] = \s[21] and \s[12] = \s[22]

    \item \emph{Unambiguous alternation}:
        \newline \ualt{\mathR[1]}{\mathR[2]} if \xfa \renv\ \xin \envD. \rL{\mathR[1]} \emath{\cap} \rL{\mathR[2]} = \sempty

    \item \emph{Unambiguous variable binding}:
        \newline \ubind{\dvarx}{\mathR[1]}{\mathR[2]} if \xfa \renv\ \xin \envD, \s[11], \s[12] \xin \rL{\mathR[1]}.
        \newline \mbox{~~~} if \s[21] \xin \rL[\exte{\renv}{\mape{\dvarx}{\s[11]}}]{\mathR[2]}, \s[22] \xin \rL[\exte{\renv}{\mape{\dvarx}{\s[12]}}]{\mathR[2]}, and \s[21] = \s[22],
        \newline \mbox{~~~} then \s[11] = \s[12]

\end{itemize}
\end{definition}

Finally, we can define what it means for a given match-reference regex, \mathR, to be strongly unambiguous in the context of a regex type environment, \rtenv:
\begin{definition}[\mathR\ is strongly unambiguous in \rtenv]
\leavevmode

\noindent
\mathR\ is strongly unambiguous in \rtenv\ if \rtenv\ is well-formed (\rtenv\ \emath{\vdash *}) and one of the following holds:

\begin{itemize}
    \item \xfa \renv\ \xin \envD,  \rL{\mathR} = \emath{\emptyset}
    \item \mathR\ = \s
    \item \mathR\ = \rstar{\mathR'},
    \uiter{\mathR'},
    and \xpr{\mathR} is strongly unambiguous in \rtenv
    \item \mathR\ = \ralt{\mathR[1]}{\mathR[2]}, \ualt{\mathR[1]}{\mathR[2]},
    and \mathR[1], \mathR[2] are strongly unambiguous in \rtenv
    \item \mathR\ = \rand{\mathR[1]}{\mathR[2]}, \uconcat{\mathR[1]}{\mathR[2]},
    and \mathR[1], \mathR[2] are strongly unambiguous in \rtenv
    \item \mathR\ = \bindin{\dvarx}{\mathR[1]}{\mathR[2]},
    \ubind{\dvarx}{\mathR[1]}{\mathR[2]}, \mathR[1] is strongly unambiguous in \rtenv, and \mathR[2] is strongly unambiguous in \extenv{\rtenv}{\dvarx}{\mathR[1]}
    \item \mathR\ = \dvarx, \rtenv\ = \xpr{\rtenv}, \xpair{\dvarx}{\xpr{\mathR}}, \xpr{\xpr{\rtenv}} and \xpr{\mathR} is strongly unambiguous in \xpr{\rtenv}
\end{itemize}

\end{definition}
Bohannon et al.\cite{Bohannon2008} provide a framework for implementing unambiguity checks for the core regular expressions. In addition, we can check that a binding expression, \bindin{\dvarx}{\mathR[1]}{\mathR[2]}, is unambiguous by verifying that \dvarx\ is used in each branch of \mathR[2]'s abstract syntax tree, and that the result of substituting \dvarx\ for \mathR[1] in \mathR[2] is unambiguous.

\subsection{Typing rules}
\label{sssec:tyrules}

\newif\ifannot\annotfalse

\begin{figure}[t]
\input{definitions/MRLens_TypingRules}
\caption{Typing Rules for Match-References Lenses}
\label{fig:mrlens-type}
\end{figure}

The typing rules are given in \autoref{fig:mrlens-type}. Again, the typing judgment is \ljudg{\xl}{\ltype{\mathR[1]}{\mathR[2]}} where \ltenv\ and \rtenv\ are type environments, \xl\ is a match-reference lens, and \mathR[1] and \mathR[2] are match-reference regular expressions. The \ltenv\ type environment maps variables to lens types and the \rtenv\ type environment maps variables to regular expressions. The \tyrule{LinkT} rule adds to the environments, and the \tyrule{VarT} uses the lens type environment, \ltenv, to look up variable lenses type. The regex type environment, \rtenv, is used to ensure that the match-reference regexes that type the match-reference lenses are well-formed. The rules are structured such that if \xl\ is well-typed in \ltenv\ and \rtenv\ then \ltenv\ and \rtenv\ are well-formed. The base cases, \tyrule{ConstT}, \tyrule{VarT}, and \tyrule{IdT}, state that the environments must be well-formed (\goodtyenvs). This condition propagates to the inductive cases, via the typing judgments that appear in their premises. This is particularly important for \tyrule{LinkT}, the one rule with a premise that modifies the type environments. Otherwise, the rules for the core bijective lenses are largely unchanged.

As with prior work, the lens types consist of strongly unambiguous regexes. This unambiguity in the lens types means that the result of applying a lens to a string is deterministic; the same substrings of a given string will always be given to the same sublenses. This deterministic parse by the lens is part of what ensures that the \ename\ lenses obey the round-tripping laws. For instance, the lens \xl\ = \xlor{\xconst{a}{A}}{\xconst{a}{B}} does not obey the round-tripping laws: \xl.put(\xl.get(B)) might yield ``A'' or ``B''. Without the unambiguity constraint, this lens would be well-typed; it would have the type: \ltype{\ralt{\rconst{a}}{\rconst{a}}}{\ralt{\rconst{A}}{\rconst{B}}}. However, given the unambiguity constraint, \xl\ is not well-typed, because the potential left-hand regex of the type, \ralt{\rconst{a}}{\rconst{a}}, parses the string ``a'' in two ways. More generally, the type system uses the unambiguity constraint to exclude lenses that would not obey the round-tripping laws.

Within this type system, \xldef{pg\_map\_MR} is well-typed. It can be typed as:
\begin{center}
   \rstar{ \bindin{\xrvar{fname1}}
        {\xrdef{pg\_fname}}
            {\rand{\rand{\xlit{<href="}}{\rand{\xrdef{pg\_url\_prefix}}{\rand{\xrvar{fname1}}{\xlit{">}}}}}
                  {\rand{\xrvar{fname1}}{\xlit{</a>}}}}}
                  
                                    \emath{\Leftrightarrow}
                                    
                                     \rstar{\bindin{\xrvar{fname2}}
                                               {\xrdef{pg\_fname}}
                                               {\rand{\rand{\xlit{[}}{\rand{\xrvar{fname2}}{\xlit{]}}}}{\rand{\xlit{(}}{\rand{\rand{\xrdef{pg\_url\_prefix}}{\xrvar{fname2}}}{\xlit{)}}}}}}
\end{center}
In this case, the variable lens \xrvar{fmap} has the type \ltype{\xrvar{fname1}}{\xrvar{fname2}}, rather than \ltype{\xrvar{fname}}{\xrvar{fname}}, because the typing rules stipulate that we must use fresh variables when typing a link lens. Nonetheless, the expressions here are alpha equivalent to the match-reference regexes we defined above: \xrdef{pg\_html\_MR} and \xrdef{pg\_md\_MR}, which represent the HTML and Markdown formats, respectively. So, given the guarantees that we get with typed bijective lenses, this means that we know that when we use \xldef{pg\_map\_MR} to translate the Project Gutenberg page, we will get valid HTML or Markdown, and the relationship between the visible and hidden file name will remain intact across translations.

\begin{figure}
        \input{definitions/MRLens_BigStepSemantics}
    \caption{Big Step Semantics for Match-Reference Lenses}
    \label{fig:lens-bigstep}
\end{figure}

\section{Match-Reference Lens Implementation}
\label{sec:mre-lens-impl}

A naive implementation of the match-reference lenses consists of a direct translation of the big-step semantics given in \autoref{fig:lens-bigstep}\footnote{A more efficient implementation is given in Musca's dissertation \cite{MuscaThesis}.}. The judgment in the operational semantics has the form: \fstep{\xl}{\s[1]}{\s[2]}, where \xl\ is the lens being applied, \emath{f} is the function (either \lget\ or \lput), \s[1] is the input string, and \s[2] is the output string. The steps take place in the context of three environments: \ltenv, the lens type environment, \rtenv, the regex type environment, and \renv, the regex value environment. The base cases, \textsc{ConstP/G}, \textsc{VarP/G}, and \textsc{IdP/G} only take a step if the environments are well coordinated (\ltenv, \rtenv, \renv\ \emath{\vdash *}), we define this property in \autoref{def:good-step-envs}. This property is propagated to the inductive cases via steps taken in the premises. 

\begin{definition}[\ltenv, \rtenv, \renv\ \emath{\vdash *}: The operational semantic environments are well-formed]
\label{def:good-step-envs}
\leavevmode 

\noindent The environments are well-formed if:
                        \begin{enumerate}
                            \item for any given lens variable, \xlvar, in the domain of \ltenv, \lu{\ltenv}{\xlvar} = \ltype{\dvarx[1]}{\dvarx[2]} if and only if \dvarx[1], \dvarx[2] are in the domains of \rtenv\ and \renv\ and
                            \item for any given regex variable, \dvarx, in the domain \rtenv\ and \renv, \lu{\renv}{\dvarx} \xin \rL{\lu{\rtenv}{\dvarx}}.
                        \end{enumerate}
\end{definition}
\noindent Essentially, this side-condition ensures that the types in the lens type environment exist in the regex type environment, and that the values in the regex value environment are in the language of their type. Note that \textsc{LinkG/P} is the only rule that updates the environments and that it does so in a way that maintains this side-condition.

Taking a look at how the steps get carried out, the simplest base case is \textsc{ConstP/G} which applies to the constant lens, \xconst{\s[1]}{\s[2]}: its \lget\ function reads \s[1] and outputs \s[2], and its \lput\ function does the inverse. The two other base cases,  \textsc{VarP/G} and \textsc{IdP/G}, first get the type of the lens, and then verify that the input string indeed belongs to that type before performing the translation. The inductive case \textsc{EStep} deals with the case where the input to the lens is an expression. In that case, the expression must first be evaluated to a string, and then the lens function gets applied to that string. Finally, taking a step (\fstep{\xl}{\s}{\s'}) in all the other inductive cases consist of three major parts: 
\begin{enumerate}
\item Infer the type of the lens, \xl, in \ltenv, \rtenv: \ljudg{\xl}{\ltype{\mathR[1]}{\mathR[2]}}
\item Use the appropriate regex (\mathR[1] if \emath{f}= \lget\ and \mathR[2] if \emath{f}=\lput)  to parse the input string \s\ into sub-strings
\item Apply \xl's sub-lenses to the corresponding sub-strings, in updated environments, in the case of \textsc{LinkG/P}.
\end{enumerate}
For instance, say we want to apply \xconcat{\xl_{1}}{\xl_{2}}'s \lget\ function to the string \s\ in the context of \ltenv, \rtenv and \renv. First, we get \xconcat{\xl_{1}}{\xl_{2}}'s type which will be a pair of concatenations: \ltype{\rand{\mathR[11]}{\mathR[12]}}{\rand{\mathR[21]}{\mathR[22]}}. With \rand{\mathR[11]}{\mathR[12]}, we then split \s\ into two strings, \rand{\s[1]}{\s[2]}, where \s[1] \xin \rL{\mathR[11]} and \s[2] \xin \rL{\mathR[12]}. Finally, we recursively apply the sub-lenses, \xl[1] and \xl[2], to their corresponding sub-strings, \s[1] and \s[2], respectively.

So, implementing the lenses involves building two main components: a type inferencer and a parser. Building the type inferencer is the relatively simple task of implementing the typing rules given in \autoref{fig:mrlens-type}. However, implementing a string parser for match-reference regexes is not so straight-forward. The denotational semantics we provided in \autoref{fig:mre-sem} do not lend themselves well to implementation; the denotation of a regex, \mathR, is a set of strings which is not necessarily finite. So, with the denotational semantics, we can only \emph{recognize} that a string, \s, is in the language of \mathR, by enumerating the members in the set \rL[\empenv]{\mathR} until \s\ is found.
Thus, to parse a string with match-reference regexes, we define a machine for \emph{deciding} string membership in their language, which we present next, in \autoref{sec:mras}.

\SetKwFunction{ToS}{ToSRE}
\SetKwFunction{ToSAcc}{ToSAcc}

\subsection{Match-reference regex automata systems}
\label{sec:mras}

\subsubsection{Introducing match-reference regex automata systems}

As the tool for parsing strings with match-reference regexes, we define a machine for deciding the language of match-reference regexes, called the \emph{\upename\ Regex Automata System} (\initename \ras), which, as discussed in \autoref{ssec:mre-vs-bref}, is inspired by past work on formalizing back-references \cite{Campeanu2009a, Campeanu2009b, Yu2004, Freydenberger2018, Schmid2019}. The name ``automata system'' comes from the fact that these machines are a coordinated collection of finite state automata (FSA). To provide an intuition for how an \initename \ras\ functions, we present an \initename \ras, \mras, that is equivalent to the  match-reference regex \bindin{\dvarx}{\rstar{\rconst{a}}}{\rand{\dvarx}{\rand{\rconst{c}}{\dvarx}}}. That is, \mras\ decides membership in the language whose strings fit the pattern: \emath{\xlit{a}^k\xlit{c}\xlit{a}^k}. 

At its core, an \initename \ras\ has a collection of $n + 1$ finite state automata, \emath{\rfsa[0], ..., \rfsa[n]}, where \rfsa[n] is the \emph{main automaton} and \rfsa[0], ..., \rfsa[n-1] are \emph{variable automata}. An \initename \ras's automata act like typical finite state automata, except steps from one state to the next may contain special symbols: \rasinvar{k}, \rasvar{k}, and \rasoutvar{k}. Briefly, \rasinvar{k} puts the variable \dvarx[k] in scope, \rasvar{k} consumes a match for \dvarx[k], and \rasoutvar{k} puts \dvarx[k] out of scope.

\begin{figure}
\begin{subfigure}{.49\textwidth}
\centering 
\includegraphics[width=.85\textwidth]{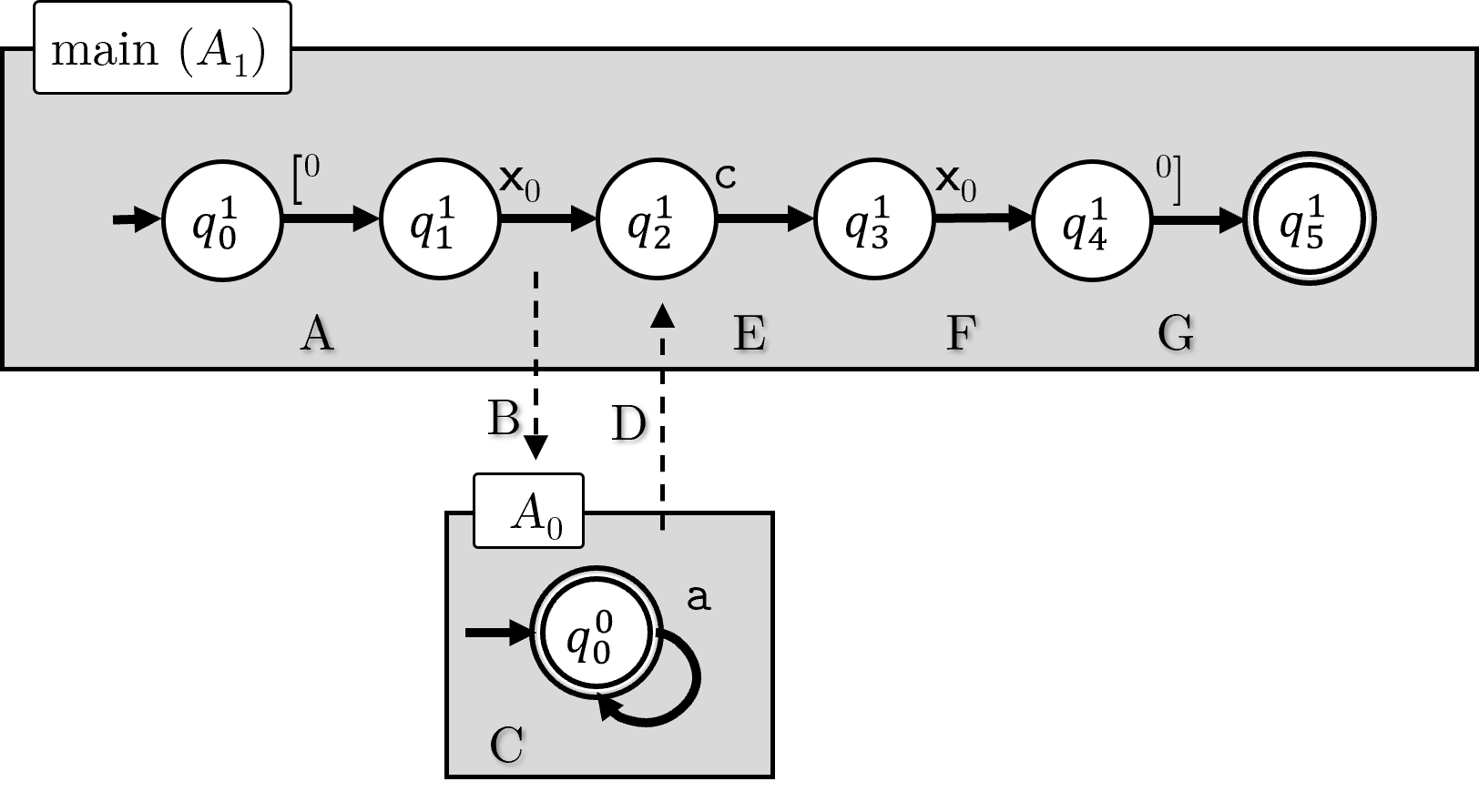}
 \vspace{1.5cm}
\caption{\emath{M}'s finite state automatas, \rfsa[0] and \rfsa[1], annotated with the steps that correspond to those taken in \autoref{fig:simp_config_steps}. The double-circled states are accept states.
}
\label{fig:mras_fsas_simp}
\end{subfigure}
\hfill
\begin{subfigure}{.49\textwidth}
\centering 
\includegraphics[width=.9\textwidth]{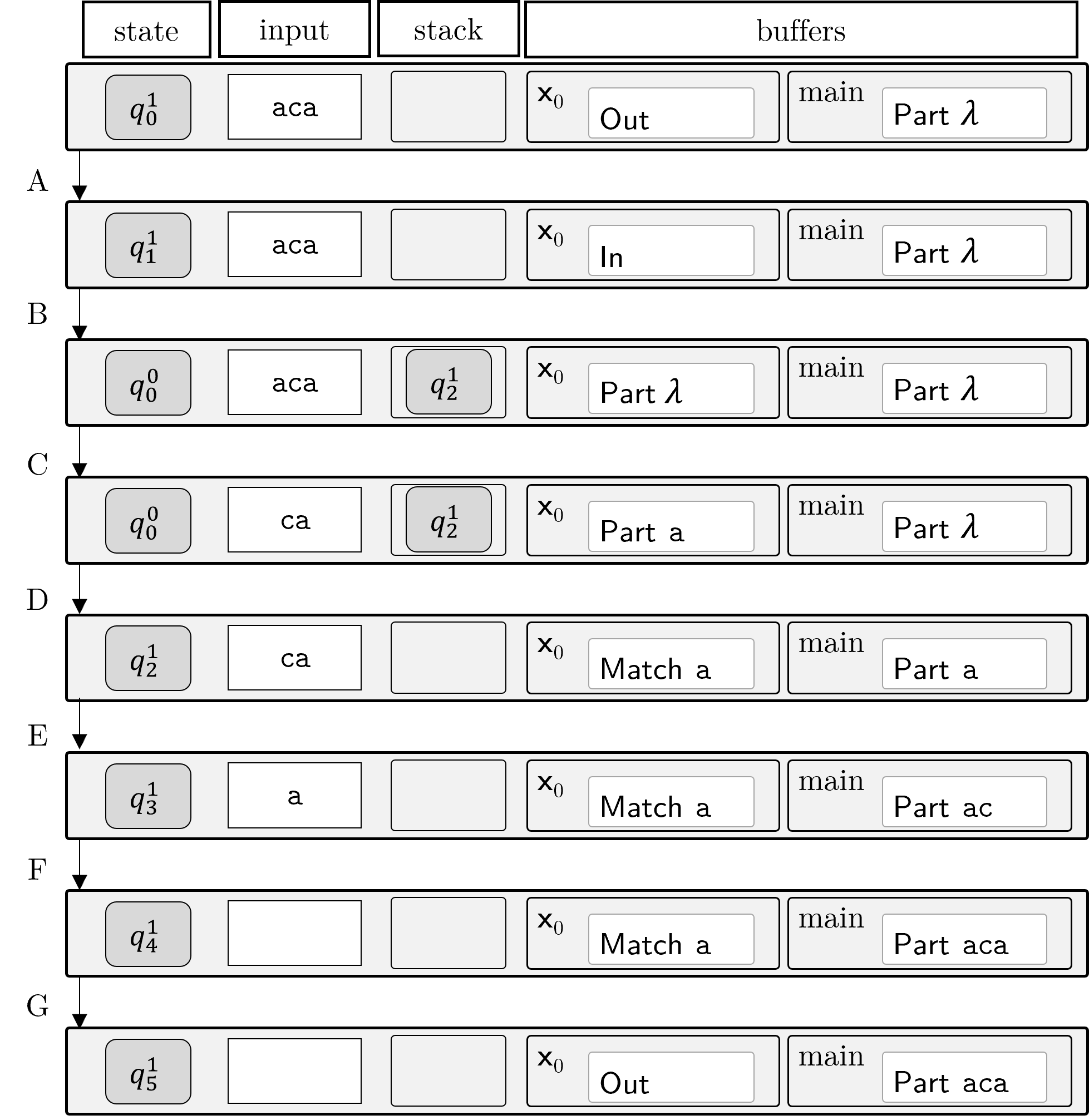}
\caption{Configurations representing the path \mras\ takes in accepting \xlit{aca}.}
\label{fig:simp_config_steps}
\end{subfigure}
\caption{An MRAS, \mras, that decides membership in the language: \emath{\xlit{a}^k\xlit{c}\xlit{a}^k}.}
\label{fig:mras_eg}
\end{figure}

In our case, \mras\ has two FSAs: \rfsa[0] and \rfsa[1] (shown in \autoref{fig:mras_fsas_simp}). \rfsa[0] is \mras's sole variable automaton, and corresponds to the variable \dvarx[0]. \rfsa[0] is a basic finite state automaton--it doesn't use special symbols--and it accepts strings in the language of \rstar{\rconst{a}}. \mras's main automaton is \rfsa[1], that is, \mras\ uses \rfsa[1] to match strings in the top-level language. Notice, \rfsa[1] \emph{does} have steps that consume the special symbols mentioned above: \rasinvar{0}, \rasvar{0}, and \rasoutvar{0}. Outside of the context of \mras, we can think of \rfsa[1] as accepting strings that match \rand{\dvarx[0]}{\rand{\rconst{c}}{\dvarx[0]}}, where \dvarx[0] is undefined. Within the context of \mras, \rfsa[1]'s special steps trigger \mras\ to change \dvarx[0]'s state or to use \rfsa[0] to find a match for \dvarx[0]. Thus, \mras\ uses \rfsa[1] in conjunction with \rfsa[0] to decide strings in the language of \bindin{\dvarx}{\rstar{\rconst{a}}}{\rand{\dvarx}{\rand{\rconst{c}}{\dvarx}}}. To illustrate how \mras\ operates, we will walk through the steps taken by \mras\ when it processes the input string \xlit{aca}. First, though, we need to introduce the concept of \initename \ras\ \emph{configurations}.

An \initename \ras\ configuration has four components: the current state, the current input, a state stack, and a set of buffers. Essentially, the state stack coordinates the automata when their operations are nested. Each buffer is associated with an automaton. The buffer associated with a variable automaton, \rfsa[k], represents the state of the variable \dvarx[k]. A variable, \dvarx[k], can be in one of four states: \vbout\ (\dvarx[k] is not in scope), \vbin\ (\dvarx[k] is in scope but matching has not begun), \vbcurr{\xstr} (\rfsa[k], so far, has matched the prefix \xstr), and \vbfound{\xstr} (\rfsa[k] found a match for \dvarx[k]: \xstr).  The buffer for the main automaton is always in the \vbcurr{\xstr} state and  essentially stores the string that has been matched so far. An \initename \ras's initial configuration on input, \xstr, starts at the main automaton's initial state, with an empty stack, and all variables out of scope. An \initename \ras\ operates by stepping from one configuration to the next, and it accepts a string, \xstr, if it can step from the initial configuration with input, \xstr, to an accepting configuration. An accepting configuration is one where all the input has been consumed, the stack is empty, and all the variables are out of scope.

\autoref{fig:simp_config_steps}, gives an example of an \initename \ras\ accepting a string, namely, \mras's configurations along a path that accepts the string \xlit{aca}. \mras\ begins in its initial configuration: the current state is \rfsa[1]'s initial state, the stack is empty and \dvarx[0] is out of scope. From this initial configuration, \mras\ takes the only step it can take, (A), and takes a step in \rfsa[1] which puts \dvarx[0] in scope, changing \dvarx[0]'s buffer's state to \vbin. \mras's next step (B) is again in \rfsa[1], which consumes \dvarx[0]. At this point, though, \dvarx[0] has no match, so \mras\ calls on \rfsa[0] to find one. The next configuration's state is \rfsa[0]'s initial state, the stack now holds the state that \mras\ will return to once a match for \dvarx[0] has been found, \rfsaq{1}{2}, and \dvarx[0]'s buffer changes to \vbcurr{\lambda}. Then, taking step (C), \mras\ consumes \xlit{a} with \rfsa[0], updating \dvarx[0]'s buffer. At this point, the configuration'
s state is a final state in \rfsa[0], so a possible match has been found for \dvarx[0], and \mras\ can return to \rfsa[1]. So, \mras\ takes step (D) going into the state that was at the top of the stack, \rfsaq{1}{2}, and changing \dvarx[0]'s buffer to \vbfound{\xlit{a}} to indicate that a possible match was found for \dvarx[0]. In step (E), \mras\ takes a step with \rfsa[1] that consumes \xlit{c}. Then, the \mras\ takes another step, (F), in \rfsa[1] that consumes \dvarx[0]. This time though, \dvarx[0] has a match, so, \mras\ does not call on \rfsa[0], rather it consumes the string in \dvarx[0]'s buffer: \xlit{a}. Finally, in (G), \mras\ takes a step in \rfsa[1], consumes \rasoutvar{0}, and puts \dvarx[0] out of scope. Thus, \mras\ has arrived at an accepting configuration: it is in \rfsa[1]'s accept state, the input string has been fully consumed, the stack is empty and \dvarx[0] is out of scope. 

\subsubsection{Match-reference regex automata systems: Formal definition}

We give the formal definition for \initename \ras\ in three parts:
\begin{itemize}
    \item \initename \ras\ automata
    \item \initename \ras\ configurations, and
    \item how an \initename \ras\ steps from one configuration to the next.
\end{itemize}

\paragraph{\initename \ras\ automata}

As mentioned above, an MRRAS includes a collection of $n + 1$ finite state automata, \rfsaset\ = \set{\rfsa[0], ..., \rfsa[n]}, where \rfsa[n] is the main automaton and \rfsa[0], ..., \rfsa[n-1] are variable automata.
Each automaton,
\rfsa[k], in \rfsaset\ has the following components: \rfsacomp[k]{k}. \rfsaqset{k} is the set of \rfsa[k]'s states;  \rasalph[k] is the automaton's alphabet (specified below);  \rfsastep{k}: (\rfsaqset{k} $\times$ \rasalph[k] $\rightarrow$ \rfsaqset{k}) is its set of transitions; \rfsainit{k} is its initial state, and \rfsafinal{k} is its set of final states. So, an \initename \ras's automata are just like typical finite state automata (aside from their augmented alphabets).

The automata each have their own alphabet: \rasalph[k]. \rasalph[k] consists of a finite input alphabet, \rasalph, extended with the set of special characters: \set{\rasvar{0}, \rasinvar{0}, \rasoutvar{0}, ..., \rasvar{k-1}, \rasinvar{k-1}, \rasoutvar{k-1}}. In the context of an \initename \ras, the special symbols have the following effect:
\begin{itemize}
    \item \rasinvar{i}: puts the variable \dvarx[i] in scope.
    \item \rasoutvar{i}: puts the variable \dvarx[i] out of scope, and in doing so, deletes the string that \dvarx[i] matched (if one was found).
    \item \rasvar{i}: if a match for \dvarx[i] exists, matches that string, otherwise, uses \rfsa[i] to find a match for \dvarx[i].
\end{itemize}
 Notice that a given automaton, \rfsa[k], can only affect the variables of automata that are ``below'' it. In particular, \rfsa[0]'s alphabet is actually just the input alphabet; none of its transitions will contain special symbols.

\paragraph{\initename \ras\ configurations}
A \initename \ras\ configuration is a tuple: \rconf{\rfsaq{i}{j}}{\xstr}{\rasstack}{\metavbuf}.
The configuration's first element, \rfsaq{i}{j}, is the \initename \ras's current state (\textit{i} is the number of the automaton, and \textit{j} is the automaton's $j^{\text{th}}$ state). That state, \rfsaq{i}{j}, belongs to the automaton \rfsa[i], which we call the \emph{active} automaton.
The  configuration's second element, \xstr, is the current input string drawn from \rstar{\rasalph[]}, not \rasalph[i]; the input string does not contain the special characters \rasinvar{k}, \rasoutvar{k} and \rasvar{k}.
The configuration's third element, \rasstack, is a stack of states, one state for each variable we are in the process of matching.
The configuration's fourth element, \metavbuf, is a collection of buffers,
\set{\vbuf{0}, ..., \vbuf{n}}, one buffer for each automaton. 

The buffer of a variable automaton, \rfsa[k], represents the state of the variable \dvarx[k]. A variable buffer is in one of four states: \vbin, \vbout, \vbcurr{\xstr}, and \vbfound{\xstr}. If \vbuf{k} = \vbout, then \metavar{k} is out of scope, otherwise, the variable is in scope. When \vbuf{k} = \vbin, the variable is in scope but has not yet appeared. When \vbuf{k} = \vbcurr{\xstr}, the first use of \metavar{k} has been encountered, the \initename \ras\ is in the process of finding a match for \metavar{k}, and the prefix of a match for \metavar{k}, \xstr, has been found. Finally, when  \vbuf{k} = \vbfound{}{\xstr}, the \initename \ras\ has found a match for \metavar{k}: \xstr. The state of the main automaton's buffer, \vbuf{n}, is always \vbcurr{\xstr}, where \xstr\ is the prefix of the input string that \initename \ras\ has consumed so far. 

An \initename \ras's starting configuration is \rconf{\rfsainit{n}}{\xstr}{\stempty}{\metavbuf[init]}, where \xstr\ is the initial input, \stempty\ is the empty state stack, and \metavbuf[init] = \set{\vbout, ..., \vbout, \vbcurr{\lambda}}. A final configuration has the form \rconf{\rfsaq{i}{k}}{ \xemp}{\stempty}{\metavbuf[accept(\xstr)]}, where \rfsaq{n}{k} \xin \rfsafinal{n} (that is, \rfsaq{n}{k} is in the main automata's set of final states) and \metavbuf[accept(\xstr)] = \set{\vbout, ..., \vbout, \vbcurr{\xstr}}.

\begin{figure}
\input{definitions/MRAS_steps}
\caption{MRRAS Steps} \label{fig:mrras-steps}
\end{figure}

\paragraph{\initename \ras\ steps}

An execution step of an MRRAS, \mras, has the form: \rconf{\rfsaq{i}{k}}{\xpref\xstr}{\rasstack}{\metavbuf} \rstep \rconf{\rfsaq{i'}{k'}}{\xstr}{\xpr{\rasstack}}{\xpr{\metavbuf}}
where \rfsaset\ is \mras's set of automata. The steps that an MRRAS can take are given in \autoref{fig:mrras-steps}. Broadly speaking, each step that \mras\ takes either consumes input (\lcmatch, \lvmatch), updates the scope of a variable (\lvin, \lvout), or changes the active automaton (\linitv, \lvret). Say the current configuration of \mras\ is \rconf{\rfsaq{i}{k}}{\xstr}{\rasstack}{\metavbuf}. In this case, the active automaton is \rfsa[i], and so, \mras\ consults both \rfsa[i]'s set of final states (\rfsafinal{i}) and \rfsa[i]'s set of transitions (\rfsastep{i}), to see what steps, if any, can be taken. 

When \rfsaq{i}{k} is in \rfsafinal{i}, this means something different depending on whether \rfsa[i] is the main automaton or a variable automaton. If \rfsa[i] is the main automaton, then, if the input is empty and all the variables are out of scope, \mras\ has reached an accepting configuration. So, \mras\ has accepted the initial input string. On the other hand, if \rfsa[i] is a variable automaton, then this means that \mras\ can take a \lvret\ step; \mras\ has found a potential match for \dvarx[i] and can return control to the automaton, \rfsa[j], that called \rfsa[i]. \mras\ updates \vbuf{i} to indicate that a match has been found for \dvarx[i], and its next state is at the top of \rasstack, \rfsaq{j}{k'}, one of \rfsa[j]'s states.

With regards to \rfsa[i]'s set of transitions, \rfsastep{i}, they might include steps that can be taken from \rfsaq{i}{k}:  \rfsastep{i}(\rfsaq{i}{k}, \emath{sym}) \emath{\rightarrow} \rfsaq{i}{k'}, where \emath{sym} \xin \rasalph[i]. Here, the steps that \mras\ can take hinge on what \emath{sym} is. If \emath{sym} is \xpref, where \xpref\ is a prefix of \xstr, then \mras\ can take a \lcmatch\ step: it consumes \xpref, updates \vbuf{i}, and enters state \rfsaq{i}{k'}. On the other hand, if \emath{sym} is \rasinvar{j}, then \mras\ can take an \lvin\ step: \mras\ puts \dvarx[j] in scope, by setting \vbuf{j} to \vbin. If \emath{sym} is \rasoutvar{j}, then \mras\ can take the \lvout\ step, putting \dvarx[j] out of scope, and deleting \dvarx[j]'s last match, if one was found. Finally, if \emath{sym} is \rasvar{j}, then \mras\ might take one of two steps: \lvmatch\ or \linitv.

More specifically, if \rfsa[i]'s set of transitions includes \rfsastep{i}(\rfsaq{i}{k}, \rasvar{j}) \emath{\rightarrow} \rfsaq{i}{k'}, then \vbuf{j} determines if \mras\ can take a \lvmatch\ or \linitv\ step. If \vbuf{j} = \vbfound{\xpref}, then \metavar{j} has already been matched, and all uses of \metavar{j} within the current scope, including this one, must continue to match the same string. So, if \xpref\ is a prefix of the current input, \lvmatch\ is applied: the active automaton remains \rfsa[i], \xpref\ is consumed from the input, and \mras's next state is \rfsaq{i}{k'}. If \vbuf{j} = \vbin\, then a match for \metavar{j} has not been found since it was put in scope, and \mras\ can take the \linitv\ step. In this case, \mras\ uses \rfsa[j] to find a match for \metavar{j}. \mras's next state is \rfsainit{j}, \rfsa[j]'s initial state, and \rfsaq{i}{k'} goes on the state stack, for when control is returned to \rfsa[i] once a match for \metavar{j} has been found.

A string, \xstr, is accepted by \mras\ if a final configuration is reachable from the initial configuration on input \xstr\ via the steps described above. 
Note that an \mras\ may be able to take several steps from a given configuration, in which case, there are branching paths in \mras's execution. However, each branch is guaranteed to terminate because \mras, on a given input, \xstr, has a finite number of configurations. The number of configurations is finite because \mras\ has a finite number of states (the union of the set of states in \mras's set of automata), the input string never grows, the stack's maximum depth is bound by the number \mras's automata, and the buffers can only store sub-strings of the input string. So, any given execution path ends in one of three ways: a stuck configuration from which no step can be taken, a previously encountered configuration, or an accepting configuration. Thus, \mras\ is guaranteed to terminate, and at that point, if no path ended in an accepting configuration, then \mras\ rejects \xstr.

\subsection{Translating match-reference regexes into MRRASs}

\subsubsection{Intermediate representation: scoped regular expressions}

We translate a match-reference regex into an MRRAS via an intermediate representation, the \emph{scoped regular expression}, SRE, that linearizes the tree structure of match-reference regexes, and pulls out a match-reference regex's nested binding expressions. This intermediate representation has an environment that maps variables to their type and has a form dedicated to marking the scope of a variable, which more closely mirrors the structure of an MRRAS' automata system when compared to a match-reference regex's tree structure.  

\paragraph{Introducing scoped regexes}
We'll illustrate scoped regexes by once again using the Projet Gutenberg hyperlinks in \autoref{fig:gut-ex-both}, focusing on the HTML format. At the core of scoped regexes are \emph{partial regexes}: extensions to regular expressions that use variables but do not define the variables' types. So, first, we construct partial regexes to represent the filename and the url prefix of the entries (these only use basic regular expression constructors):
\begin{align*}
\xrdef{pg\_fname\_r} &:= \xlit{GUTINDEX.}[\xlit{0-9}][\xlit{0-9}][\xlit{0-9}][\xlit{0-9}] \\
\xrdef{pg\_url\_prefix\_r} & := \xlit{https://www.gutenberg.org/dirs/} 
\end{align*} 
Next, we write partial regexes that represent a single line of the hyperlinks in HTML. To do so, we make use of the two forms that extend simple regexes: variables, \dvarx, and scoping regexes, \rscope{\rpart}{\dvarx}. The scoping regex establishes that \dvarx\ is in scope in \rpart, meaning that \dvarx\ may validly appear in \rpart. The partial regex for a single line of the hyperlinks in HTML is:
\begin{align*}
    \xrdef{pg\_html\_line\_r} &:= \rscope{\rand{\rand{\xlit{<href="}}{\rand{\xrdef{pg\_url\_prefix\_r}}{\rand{\xrvar{fname}}{\xlit{">}}}}}
                          {\rand{\xrvar{fname}}{\xlit{</a>}}}}{\xrvar{fname}}
\end{align*}
As a partial regex, \xrdef{pg\_html\_line\_r} indicates where \xrvar{fname} is in scope and the position of matches for the variable \xrvar{fname}, but it doesn't specify \emph{what} \xrvar{fname} will match. 

Finally, we construct a scoped regex that represent lists of hyperlinks in HTML, making use of the partial regexes we defined above. Scoped regexes consist of a \emph{main} partial regex, \rpart, which is the body of the scoped regex, and a \emph{definition environment}, \sd, that maps variables to their types: partial regexes. Scoped regexes pair these two components in the form: \sre{\rpart}{\sd}. So, the scoped regex for a list of Project Gutenberg HTML hyperlinks is:
\begin{align*}
    \xrdef{pg\_html\_SR} &:= \sre{\rstar{\xrdef{pg\_html\_line\_r}}}{[\xpair{\xrvar{fname}}{\xrdef{pg\_fname\_r}}]}
\end{align*}
In the context of a scoped regex, \xrvar{fname} is now given a type: \xrdef{pg\_fname\_r}. So, all instances of \xrvar{fname} in \xrdef{pg\_html\_line\_r} match the same string, where that string is in the language of \xrdef{pg\_fname\_r}. Also, \xrdef{pg\_html\_line\_r} is starred in \xrdef{pg\_html\_SR}, which means that the match for \xrvar{fname} is reset each time \xrdef{pg\_html\_line\_r} is iterated, as \xrvar{fname} leaves and re-enters scope. 

\paragraph{Scoped regexes: formal definitions}
\begin{figure}
    \input{definitions/SR_denotational_simple}
    \caption{Scoped regular expression denotational semantics}
    \label{fig:sre-semantics}
\end{figure}

\begin{figure}
\centering
\begin{tabular}{|c|l|}
\hline
\subd{\sd}{\dvarx}
 & \subd{\sd}{\dvarx} = \xpr{\sd}
     where: \sd\ = \xpr{\sd}, \xpair{\dvarx}{\rpart}, \xpr{\sd'} \\ \hline
\dvarx[i]
 & The variable, \dvarx, at index \emath{i}, where \emath{i} = \numd{\subd{\sd}{\dvarx}} \\ \hline
\subd{\sd}{i}
 &  Shorthand for \subd{\sd}{\dvarx[i]} \\ \hline
\rpart[i]
 &  The partial regex that is \dvarx[i]'s type, that is: \rpart[i] = \lu{\sd}{\dvarx[i]}  \\
\hline
\end{tabular}

\caption{Notation for referring to elements of a scoped regex's definition environment, \sd}
\label{tab:sd-notation}
\end{figure}
As mentioned above, scoped regexes pair a partial regex, \rpart, with a definition environment, \sd, that maps variables to partial regexes. Formally, the syntax of a scoped regex is:

\ginit{\rpart}   
    \rconst{c} 
    \gsep\ \dvarx\
    \gsep\ \rstar{\rpart} 
    \gsep\ \rand{\rpart}{\rpart} 
    \gsep\ \ralt{\rpart}{\rpart}
    \gsep\ \rscope{\rpart}{\dvarx}

\ginit{\sd} \empsd \gsep \extsd{\sd}{\dvarx}{\rpart}

\ginit{SRE} \sre{\rpart}{\sd}

\sd\ is an ordered environment, so we can refer to entries by their index. \autoref{tab:sd-notation} introduces notation that uses the index of entries to refer to elements of \sd. We will assume, for simplicity, that each variable is unique the domain of \sd. The index of the left-most element in \sd\ is 0 and the index of the right-most element in \sd\ is \numd{\sd} - 1.

The denotational semantics for SREs are given in \autoref{fig:sre-semantics}. The semantics are evaluated in the context of the value environment, \renv, that maps variables to strings. 
Again, the semantics of the basic regular expression constructors are standard and don't make explicit use of either \renv\ or \sd. Partial regexes extend the core regular expressions with two expressions: the variable form, \dvarx, and, the scoping form, \rscope{\rpart}{\dvarx}. The SRE's variable form, \dvarx, has essentially the same semantics as that of the match-reference regex's: it denotes the set with a single string: \lu{\renv}{\dvarx}. The scoping expression, \rscope{\rpart}{\dvarx}, limits the scope of \dvarx\ to the body of \rpart\ and the type of \dvarx\ is defined in \sd. So, the denotation of \rscope{\rpart}{\dvarx} in \renv\ is the language of \rpart\ in the set of value environments that consist of \renv\ extended with all the strings in the language of \rL{\lu{\sd}{\dvarx}}.

\subsubsection{Translations}
\paragraph{Match-reference regex to scoped regex}

\begin{algorithm}[t]
\Fn{\ToS{\mathR}}{
\Fn{\ToSAcc{\mathR, \sd}}
{
    \KwMatch{\mathR}
    {
        \lPat{\rconst{c}}
            {\Return{\sre{\rconst{c}}{\sd}}}
        \lPat{\dvarx}
            {\Return{\sre{\dvarx}{\sd}}}
        \Pat{\rstar{\xpr{\mathR}}}
            {\sre{\rpart}{(\sd, \sd')} \emath{\leftarrow} \ToSAcc{\xpr{\mathR}, \sd} \;
            \Return{\sre{\rstar{\rpart}}{(\sd, \sd')}}}
        \Pat{{\mathR[1]} \emath{\xgenop} {\mathR[2]}}
            {\sre{\rpart[1]}{(\sd, \sd_1)} \emath{\leftarrow} \ToSAcc{\mathR[1], \sd} \;
            \sre{\rpart[2]}{(\sd, \sd_2)} \emath{\leftarrow} \ToSAcc{\mathR[2], \sd} \;
            \Return{\sre{{\rpart[1]} \xgenop\ {\rpart[2]}}{(\sd, \sd_1, \sd_2)}}}
        \Pat{\bindin{\dvarx}{\mathR[1]}{\mathR[2]}}
            {\sre{\rpart[1]}{(\sd, \sd_1)} \emath{\leftarrow} \ToSAcc{\mathR[1], \sd} \;
            \sre{\rpart[2]}{(\sd, \sd_1, \xpair{\dvarx}{\rpart[1]}, \sd_2)}
                \emath{\leftarrow}
                \ToSAcc{\mathR[2], \emath{(\sd, \sd_1, \xpair{\dvarx}{\rpart[1]})}} \;
           \Return{\sre{\rscope{\rpart[2]}{\dvarx}}{(\sd, \sd_1, \xpair{\dvarx}{\rpart[1]}, \sd_2)}}}
    }
    }
    \Return{\ToSAcc{\mathR, \empsd}}
}
\caption{Converting MREs to SREs}
\label{alg:mretosre}
\end{algorithm}

\autoref{alg:mretosre} translates a match-reference regex, \mathR, into an equivalent scoped regex, \sre{\rpart}{\sd}. In essence, the body of \mathR\ becomes the main regex \rpart, and all the defined variable types get extracted and put in \sd. We assume that defined variables are unique in a match-reference regex, given that for any match-reference regex with non-unique variable names, we can define an alpha-equivalent one with unique variable names. This in turn means that the variables defined in the definition environment of the resulting SRE will have unique variables. 

The translation from match-reference regex to SRE takes the tree structure of match-reference regexes and flattens it into a linear structure in the definition environment of the scoped regex. However, the equivalence between a match-reference regex and the scoped regex it is translated to is maintained, because there is an implicit tree structure that remains in the resulting scoped regex as shown by Musca \cite{MuscaThesis}. So, the languages of a match-reference and its scoped regex are equivalent. 

\paragraph{SRE to MRRAS}

\SetKwFunction{ToMRRAS}{to-MRRAS}
\SetKwFunction{XtoIndex}{var-to-ind}
\SetKwFunction{ToFSAs}{to-FSA-Sys}
\begin{algorithm}[t]
    \Fn{\ToMRRAS{\sre{\rpart}{\sd}}} 
        {    \Fn(\tcp*[f]{convert definitions to variable automata}){\ToFSAs{\sd}}
                {\KwMatch{\sd}
                    {\lPat{\empsd}{\empsd}
                     \lPat{\xpr{\sd}, \xpair{\dvarx[i]}{\rpart}}{\ToFSAs{\xpr{\sd}}, \rtofsa{\rpart}{i}}}}
         \xpair{\rpart^{I}}{\sd^{I}} \emath{\leftarrow} \XtoIndex{\sre{\rpart}{\sd}} \tcp*[f]{convert variables to indices} \;
         \rfsaset[var] \emath{\leftarrow } \ToFSAs{\emath{\sd^{I}}} \tcp*[f]{get variable automata} \;
         \rfsa[main] \emath{\leftarrow} \rtofsa{\rpart^{I}}{\numd{\sd}} \tcp*[f]{get main automaton} \;
        \Return{(\rfsaset[var], \rfsa[main])}\;
        }
\caption{Converting SREs to MRRASs automata system}
\label{alg:sre-to-mras}
\end{algorithm}

\begin{figure}[t]
    \begin{center}
    \input{definitions/rpart-to-fsa}
    \end{center}
    \caption{Constructing an MRRAS FSA from a partial regex}
    \label{fig:rpart-to-fsa}
\end{figure}

The high-level algorithm for translating an SRE, \sre{\rpart}{\sd}, into an MRRAS's automata system, \rfsaset, is given in \autoref{alg:sre-to-mras}. The first step is re-encoding the variables in \sre{\rpart}{\sd} such that the variables are solely referred to by their index in \sd, where the index of a variable is that given in \autoref{tab:sd-notation}. We call the resulting expression: \sre{\rpart^{I}}{\sd^{I}}. We then translate the partial regexes into automata using the formula represented in \autoref{fig:rpart-to-fsa} (further discussed below). The \emath{k^{th}} partial regex in \emath{\sd^{I}}, \rpart[k], gets translated into the \emath{k^{th}} automaton, \rfsa[k], in \rfsaset, representing the type for the \emath{k^{th}} variable \dvarx[k]. The main regex, \emath{\rpart^{I}}, gets translated into the main automaton in \rfsaset, \rfsa[\numd{\sd}].

\autoref{fig:rpart-to-fsa} details the translation of the \emath{i^{th}} partial regex, \rpart[i] into the the \emath{i^{th}} MRRAS automaton, \rfsa[i]. First, the alphabet for \rfsa[i] augments the input alphabet, \rasalph, with the special symbols, \rasvar{k}, \rasinvar{k} and \rasoutvar{k}, where \emath{k} goes up to but does not include the index of the current automata, \emath{i}. This means that transitions in this automaton can only affect the status of variables smaller than it. The other components of \rfsa[i] are built out of \rpart[i] recursively. The core regex expressions are translated into FSAs in the usual manner. So, we focus on the translation of the new expressions: the variable and scoping expressions.

Variable expressions are a base case, like the constant expression. Two new states make up the set of states: \rfsaq{i}{j^{\dag}} and \rfsaq{i}{k^{\dag}}. They are initial state and final state, respectively. The only step that this machine can take is a variable step from the initial to final state: \fd{\rfsaq{i}{j^{\dag}}}{\rasvar{h}}{\rfsaq{i}{k^{\dag}}}.
For the scoping expression, \rscope{\rpart'}{\dvarx[h]}, we create a machine that opens the scope of \dvarx[h] before matching \xpr{\rpart} and closes the scope of \dvarx[h] once the match has been found. First, we get the FSA for \xpr{\rpart}, where, \rtofsa{\xpr{\rpart}}{i} = \rfsacomp[\xpr{\rpart}]{i}. We augment \rfsaqset{\xpr{\rpart}} with two new states: \rfsaq{i}{j^{\dag}} and \rfsaq{i}{k^{\dag}}. The new initial state becomes \rfsaq{i}{j^{\dag}} and we add a transition that opens the scope for \dvarx[h] going from our new initial state to the previous initial state \rfsainit{\rpart'}: \fd{\rfsaq{i}{j^{\dag}}}{\rasinvar{h}}{\rfsainit{\xpr{\rpart}}}. We also make \rfsaq{i}{k^{\dag}} the sole final state, and add a set of  transitions that close the scope of \dvarx[h], going from the previous final states to this new final state: \defset{\fd{q}{\rasoutvar{h}}{\rfsaq{i}{k^{\dag}}}}{q \xin \rfsafinal{\xpr{\rpart}}}.

So, we translate \sre{\rpart}{\sd} into a set of automata, \rfsaset, and pair each automaton in \rfsaset\ with a variable buffer. We then run the resulting MRRAS on a string, \s, as detailed in \autoref{sec:mras}, to decide whether \s\ is in the language of \sre{\rpart}{\sd}.

\subsubsection{A note on why the translation works}
At first glance, the semantics of match-references regexes and their \initename \ras s have the potential to diverge: the semantics of match-reference regexes (given in \autoref{fig:MRL-DenSem}) and the operation of \initename \ras's (given in \autoref{fig:mrras-steps}) differ in when the match for a variable is found. For match-reference regexes, a variable's match is set at the point at which it enters into scope. On the other hand, \initename \ras's find matches for a variable when that variable is first encountered in the body an expression. In general, for \initename \ras s, variables \emph{could} match a different set of strings depending on when their matches are found: additional variables may come into scope between when a variable itself comes into scope and when that variable occurs in the body of an expression. However, the structure of a well-formed match-reference regex (which is embedded into the structure of its \initename \ras) prevents any such divergence.

The structure of well-formed match-reference regexes, and thus, the structure of their corresponding \initename \ras's, means the variables in scope are the same for both the expression that puts a variable in scope and that variable's definition, as we illustrate in the following example:
\begin{example}[\mathR\ = \bindin{\dvarx[0]}{\rstar{\rconst{a}}}{\bindin{\dvarx[1]}{\rstar{\rconst{b}}}{\rand{\rand{\dvarx[0]}{\dvarx[1]}}{\rand{\dvarx[0]}{\dvarx[1]}}}}] 
In \mathR, when \dvarx[0] comes into scope, in the expression \bindin{\dvarx[1]}{\rstar{\rconst{b}}}{\rand{\rand{\dvarx[0]}{\dvarx[1]}}{\rand{\dvarx[0]}{\dvarx[1]}}}, no other variables are in scope but, at the location where \dvarx[0] first appears, \dvarx[1] is also in scope. However, when \dvarx[0]'s definition is declared in \mathR, no other variables are in scope. So, \dvarx[0]'s definition cannot make a valid use of \dvarx[1], that is, \dvarx[1] cannot appear free in \dvarx[0]'s definition. 
\end{example}
\noindent In other words, the variables that can appear free in a variable \dvarx's definition are exactly those that are in scope when \dvarx\ first enters into scope. This fact still holds when well-formed match-reference regexes are translated into \initename \ras s. So, despite differing in when a variable's match is set, the languages of a match-reference regex and its corresponding \initename \ras\ are equivalent.

\subsection{Summary}
To sum up, when applying a lens, \xl, to a string, \s, we can first infer the type of \xl\ with an implementation of the typing rules given in \autoref{fig:mrlens-type}, giving us \ljudg{\xl}{\ltype{\mathR[1]}{\mathR[2]}}. Then, depending on whether we are applying the \lget\ or \lput\ function of \xl, we convert \mathR[1] or \mathR[2] into an MRRAS with \autoref{alg:mretosre} and \autoref{alg:sre-to-mras}. We use that MRRAS to parse \s\ into sub-strings to which we can apply sub-lenses of \xl, according to the big-step semantics given in \autoref{fig:lens-bigstep}.

\section{Conclusion}
Programmers often need to write bidirectional translations, and domain-specific languages that implement lenses simplify the process of writing such translations. In particular, our work builds on a DSL \cite{Bohannon2008, Foster2009} that provides strong correctness guarantees, ensuring that the translations behave as expected. These strong guarantees are provided for formats characterized by regular expressions, however, which excludes formats with internal dependencies. Our work extends the core bijective lenses to allow for formats that have dependencies represented by a common extension to regular expressions: PCRE back-references. 

To that end, we have designed two related languages: \emph{match-reference regular expressions} and \emph{match-reference bijective lenses}. Match-reference regular expressions are a theoretically well-defined version of PCRE back-references. Match-reference bijective lenses extend a set of typed bijective lenses and allow users to specify translations between a richer set of languages than the core bijective lenses. The match-reference regular expressions act as types for the match-reference bijective lenses, and allow us to provide strong guarantees about the correctness of match-reference lens transformations. We also provide a template for implementing the match-reference bijective lenses based on a big-step semantics and a machine for deciding string membership in the language of match-reference regexes.
 
\paragraph*{Acknowledgements}
This research was developed with funding from the Defense Advanced Research Projects Agency (DARPA) under the SafeDocs program (award number AWD1006437). The views, opinions and/or findings expressed are those of the authors and should not be interpreted as representing the official views or policies of the Department of Defense or the U.S. Government.

\printbibliography

\end{document}

%% file: packages.tex
\usepackage{amsmath}                
\usepackage{amsthm}                 
\usepackage{subcaption}             
\usepackage{aliascnt}
\usepackage{amssymb}
\usepackage{mathpartir}             
\usepackage{stmaryrd}               
\usepackage{framed}                 
\usepackage[boxed, noline]{algorithm2e}
\usepackage{graphicx}
\usepackage{enumitem}               
\usepackage{xcolor}                  
\usepackage{hyperref}               
\usepackage[doi=false,url=false,isbn=false]{biblatex}

%% file: macros/general_formatting.tex
\setlist{nosep} 
    
    \newcommand{\emath}[1]{\ensuremath{#1}}

    \newcommand{\ginit}[1]{\ensuremath{#1 := }}
    \newcommand{\gsep}{\ensuremath{|}}

    \newcommand{\xlit}[1]{\texttt{#1}}
    \newcommand{\xrdef}[1]{\textbf{\textsf{#1}}}
    \newcommand{\xldef}[1]{\textit{\textsf{{#1}}}}
    
    \newcommand{\xrvar}[1]{\textit{\textbf{#1}}}

\DontPrintSemicolon
\SetAlgoNoLine
\SetFuncSty{textsc}
\SetKwIF{Pat}{Do}{Wild}{\emath{|}}{\emath{\rightarrow}}{}{}{}

\SetKwFor{KwMatch}{match}{with}{}
\SetKwProg{Fn}{function}{}{}
    
    \newcommand{\xdot}{\ensuremath{\cdot}}              

    \newcommand{\xpr}[1]{\ensuremath{#1'}}    

    \newcommand{\xthen}{ \ensuremath{\rightarrow} }
    \newcommand{\xfa}{\ensuremath{\forall}}
    \newcommand{\xand}{\ensuremath{\wedge} }

    \newcommand{\sempty}{\ensuremath{\emptyset}}
    \newcommand{\set}[1]{\{\ensuremath{#1}\}}
    \newcommand{\defset}[2]{\set{#1 \ | \ #2}}
    \newcommand{\xpair}[2]{\ensuremath{(#1 , \ #2)}}

    \newcommand{\xunion}{\ensuremath{\cup} }
    \newcommand{\xin}{\ensuremath{\in} }
    \newcommand{\xnin}{\ensuremath{\notin} }

    \newcommand{\mathL}[1][]{\ensuremath{\mathcal{L}_{#1}}}
    \newcommand{\mathR}[1][]{\ensuremath{\mathcal{R}_{#1}}}

    \newcommand{\empenv}{\ensuremath{\bullet}}
    \newcommand{\renv}[1][]{\ensuremath{\rho_{#1}}}             
    \newcommand{\lenv}[1][]{\ensuremath{\sigma_{#1}}}           
    \newcommand{\rtenv}[1][]{\ensuremath{\Gamma_{#1}}}          
    \newcommand{\ltenv}[1][]{\ensuremath{\Delta_{#1}}}          

	\newcommand{\envD}[1][\rtenv]{\ensuremath{\llbracket #1 \rrbracket}}    

    \newcommand{\lu}[2]{\ensuremath{#1(#2)}}                    
    \newcommand{\xdom}[1]{\ensuremath{dom(#1)}}                 
    \newcommand{\extenv}[3]{\ensuremath{#1\{#2 \mapsto #3\}}}   
    \newcommand{\mape}[2]{\ensuremath{#1 \mapsto #2}}           
    \newcommand{\exte}[2]{\ensuremath{#1\{#2\}}}                

   \newcommand{\tta}{\texttt{a}}
   \newcommand{\ttb}{\texttt{b}}
   \newcommand{\ttc}{\texttt{c}} 
   
   \newcommand{\cbox}[2]{{%
     \setlength{\fboxsep}{.2ex}
     \fcolorbox{#1}{white}{\strut#2}%
   }}
   \newcommand{\wcbox}[2]{{%
     \setlength{\fboxsep}{.4ex}
     \fcolorbox{#1}{white}{\strut#2}%
   }}
   
    \newcommand{\cunderline}[2]{\text{\color{#1}{\underline{\vphantom{g}\color{black}#2}}\color{black}}}   

%% file: macros/match-reference.tex
\newcommand{\rconst}[1]{\ensuremath{\text{#1}}}
\newcommand{\rstar}[1]{\ensuremath{{#1}^{*}}}
\newcommand{\rand}[2]{\ensuremath{#1 \cdot #2}}
\newcommand{\ralt}[2]{\ensuremath{#1 + #2}}

\newcommand{\bindin}[3]{\ensuremath{(#1:#2, #3)}}
\newcommand{\dvar}[1][x]{\ensuremath{#1}}
\newcommand{\dvarx}[1][]{\ensuremath{x_{#1}}}

\newcommand{\uconcat}[2]{\ensuremath{#1 \cdot^{!_{\rtenv}} #2}}
\newcommand{\ualt}[2]{\ensuremath{#1 +^{!_{\rtenv}} #2}}
\newcommand{\uiter}[1]{\ensuremath{#1^{*^{!_{\rtenv}}}}}
\newcommand{\ubind}[3]{\ensuremath{(#1 :^{!_{\rtenv}} #2, #3)}}

    \newcommand{\fvs}[1]{\ensuremath{\textbf{fv}(#1)}}          

    \newcommand{\s}[1][]{\ensuremath{{{s}_{#1}}}}               
    \newcommand{\rL}[2][\renv]{\ensuremath{{\llbracket #2 \rrbracket}_{#1}}}            



%% file: macros/mras.tex
\newcommand{\mras}[1][]{\ensuremath{M_{#1}}}
\newcommand{\rasalph}[1][]{\ensuremath{\mathrm{\Sigma}_{#1}}}
\newcommand{\rfsa}[1][]{\ensuremath{A_{#1}}}
\newcommand{\rfsaset}[1][]{\ensuremath{{\mathcal{S}_{#1}}}}

\newcommand{\rconf}[4]{(#1, #2, #3, #4)} 
 
\newcommand{\rfsaq}[2]{\ensuremath{{q^{#1}_{#2}}}}
\newcommand{\rfsaqset}[1]{\ensuremath{Q_{#1}}}

\newcommand{\rfsastep}[1]{\ensuremath{\delta_{#1}}}

\newcommand{\rfsainit}[1]{\ensuremath{q_{I_{#1}}}}
\newcommand{\rfsafinal}[1]{\ensuremath{Q_{F_{#1}}}}
\newcommand{\rasstack}[1][]{\ensuremath{\mathrm{\Gamma}_{#1}}}
\newcommand{\rfsacomp}[2][]{\ensuremath{(\rfsaqset{#1}, \rasalph[#2], \rfsastep{#1}, \rfsainit{#1}, \rfsafinal{#1})}}

\newcommand{\stempty}{\ensuremath{\cdot}}
\newcommand{\stpush}[2]{\ensuremath{\textnormal{\textsf{push(}}#1, #2\textnormal{\textsf{)}}}}
\newcommand{\sttop}[1]{\ensuremath{\textnormal{\textsf{top(}}#1\textnormal{\textsf{)}}}}
\newcommand{\stpop}[1]{\ensuremath{\textnormal{\textsf{pop(}}#1\textnormal{\textsf{)}}}}

\newcommand{\rasvar}[1]{\ensuremath{\textbf{\textsf{x}}_{#1}}}
\newcommand{\rasinvar}[1]{\ensuremath{\textbf{\textsf{[}}^{#1}}}
\newcommand{\rasoutvar}[1]{\ensuremath{{}^{#1}\textbf{\textsf{]}}}}   

\newcommand{\metavar}[1]{\ensuremath{x_{#1}}}
\newcommand{\metavbuf}[1][]{\ensuremath{\mathcal{V}_{#1}}}
\newcommand{\vbuf}[1]{\ensuremath{V_{#1}}}
\newcommand{\vbin}{\textsf{In}}
\newcommand{\vbout}{\textsf{Out}}
\newcommand{\vbcurr}[1]{\ensuremath{\textsf{Part } #1}}
\newcommand{\vbfound}[1]{\ensuremath{\textsf{Match } #1}}

\newcommand{\xstr}[1][]{\ensuremath{\omega_{#1}}}
\newcommand{\xpref}[1][]{\ensuremath{\alpha_{#1}}}
\newcommand{\xemp}{\ensuremath{\lambda}}

\newcommand{\rtofsa}[2]{\textbf{\textsf{to-fsa}}\ensuremath{(#1, #2)}}

\newcommand{\fd}[3]{(#1,#2,#3)}

\newcommand{\rstep}[1][\rfsaset]{\ensuremath{~{}_{#1}{\mapsto}}~}  

\newcommand{\getfsad}[1]{\ensuremath{#1.\delta}}
\newcommand{\getfsaqf}[1]{\ensuremath{#1.{Q_{f}}}}

\newcommand{\lcmatch}{\textsc{CharMatch}}
\newcommand{\lvmatch}{\textsc{VarMatch}}
\newcommand{\lvin}{\textsc{SetVarIn}}
\newcommand{\lvout}{\textsc{SetVarOut}}
\newcommand{\linitv}{\textsc{InitVar}}
\newcommand{\lvret}{\textsc{VarRet}}

%% file: macros/scoped_regex.tex
\newcommand{\sre}[2]{\ensuremath{\langle #1; #2 \rangle}}

\newcommand{\rpart}[1][]{\ensuremath{r_{#1}}}
\newcommand{\sd}{\ensuremath{D}}
\newcommand{\rscope}[2]{\ensuremath{[^{#2}#1]}}

\newcommand{\numd}[1]{\ensuremath{|#1|}}
\newcommand{\subd}[2]{\ensuremath{#1|_{#2}}}
\newcommand{\extsd}[3]{\ensuremath{#1, \xpair{#2}{#3}}}
\newcommand{\empsd}{\emath{[~]}}



%% file: macros/mrlens.tex
    \newcommand{\xl}[1][]{\ensuremath{{l_{#1}}}}                      

    \newcommand{\stepenvs}{\ltenv, \rtenv, \renv}
    \newcommand{\goodstepenvs}[1][\stepenvs]{\ensuremath{#1 \vdash \ast}}

    \newcommand{\lget}[1][\stepenvs]{\ensuremath{\text{\textsf{get}}_{}}}            
    \newcommand{\lput}[1][\stepenvs]{\ensuremath{\text{\textsf{put}}_{}}}            
                              \newcommand{\lgpf}[3][\stepenvs]
                                                    {\ensuremath{\langle #2.f(#3) | #1 \rangle }}

    \newcommand{\fstep}[4][\stepenvs]{\ensuremath{\lgpf[#1]{#2}{#3} \Downarrow #4}}

    \newcommand{\lenskw}[1]{\textsc{#1}}
    \newcommand{\xconst}[2]{\lenskw{const}\ensuremath{({#1}, {#2})}}
    \newcommand{\xiter}[1]{\lenskw{iter}\ensuremath{({#1})}}
    \newcommand{\xconcat}[2]{\lenskw{concat}\ensuremath{({#1}, {#2})}}
    \newcommand{\xswap}[2]{\lenskw{swap}\ensuremath{({#1}, {#2})}}
    \newcommand{\xlor}[2]{\lenskw{or}\ensuremath{({#1}, {#2})}}
    \newcommand{\xcomp}[2]{\ensuremath{({#1}; {#2})}}
    \newcommand{\xid}[1]{\lenskw{id}\ensuremath{({#1})}}

    \newcommand{\xlab}[3]{\lenskw{link}(\ensuremath{{#1}: {#2}, {#3}})}
    \newcommand{\xlvar}[1][]{\ensuremath{y_{#1}}}

    \newcommand{\denotes}[2][\lenv]{\ensuremath{\llbracket #2 \rrbracket_{#1}}}

    \newcommand{\goodtyenvs}[1][\tyenvs]{\ensuremath{#1 \vdash \ast}}
    \newcommand{\ltype}[2]{\ensuremath{#1 \Leftrightarrow #2}}      
    \newcommand{\ljudg}[3][\tyenvs]{\ensuremath{#1 \vdash #2 : #3}}      

    \newif\ifannot\annottrue    
     
    \newcommand{\ANNOT}[2]{\ifannot\ensuremath{\set{#1, #2}}\fi}

\makeatletter
\newcommand{\gpt}{%
  \ensuremath{\mathrel{%
    \vbox{\offinterlineskip\m@th
      \ialign{%
        \hfil##\hfil\cr
        $\scriptscriptstyle\lget\mathstrut$\cr
        \noalign{\vspace{0ex}}
        \vtop{%
          \ialign{%
            \hfil##\hfil\cr
            $\rightleftharpoons$\cr
            $\scriptscriptstyle\lput\mathstrut$\cr
          }%
        }\cr
      }%
    }%
  }}%
}
\makeatother

%% file: macros/colors.tex
\definecolor{dkblue}{rgb}{0,0.1,0.5}
\definecolor{dkcyan}{rgb}{0.1, 0.3, 0.3}
\definecolor{dkgreen}{rgb}{0,0.3,0}
\definecolor{dkred}{rgb}{0.6,0,0}
\definecolor{dkpurple}{rgb}{0.7,0,0.4}
\definecolor{olive}{rgb}{0.4, 0.4, 0.0}
\definecolor{orange}{rgb}{0.9,0.6,0.2}

\definecolor{lightyellow}{RGB}{255, 255, 179}
\definecolor{lightgreen}{RGB}{170, 255, 220}
\definecolor{teal}{RGB}{141,211,199}
\definecolor{darkbrown}{RGB}{121,37,0}
\definecolor{princetonorange}{RGB}{255,143,0}
\definecolor{tmlblue}{RGB}{0,58,120}       



%% file: definitions/MRE_grammar.tex
\ginit{\mathR} 
    \rconst{c} \gsep\ 
    \dvarx\ \gsep\
    \rstar{\mathR} \gsep\ 
    \ralt{\mathR}{\mathR} \gsep\
    \rand{\mathR}{\mathR} \gsep\ 
    \bindin{\dvarx}{\mathR}{\mathR}

 where: \rconst{c} \xin \emath{\Sigma}, \dvarx\ \xin \emath{V}, and \emath{\Sigma \cap V = \emptyset}

%% file: definitions/MRE_Semantics_Simple.tex
\begin{align*}
    \rL{\rconst{c}}
        & = \set{\rconst{c}} \\
    \rL{\dvar} & =  \set{\lu{\renv}{\dvarx}} \\        
    \rL{\rstar{\mathR}} 
        & = \rstar{\rL{\mathR}} \\
    \rL{\ralt{\mathR[1]}{\mathR[2]}}
       & = 
          \rL{\mathR[1]}  \xunion \rL{\mathR[2]}     
        \\
    \rL{\rand{\mathR[1]}{\mathR[2]}}
        & = \defset{\s[1] \xdot \s[2]}
                   {\s[1] \xin \rL{\mathR[1]}
                      \xand \s[2] \xin \rL{\mathR[2]}} 	\\
    \rL{\bindin{x}{\mathR[1]}{\mathR[2]}}
                        & = \defset{\s}
                                   {\xpr{\s} \xin \rL{\mathR[1]} \xand
                                   \s \xin \rL[{\renv \{\dvar \mapsto \xpr{\s}}\}]{\mathR[2]}} 
\end{align*}

%% file: definitions/MRLens_Grammar.tex
\ginit{\xl}
    \xconst{\s}{\s}
    \gsep\ \xid{\mathR}
    \gsep\ \xiter{\xl}
    \gsep\ \xconcat{\xl}{\xl}
    \gsep\ \xswap{\xl}{\xl}
    \gsep\ \xlor{\xl}{\xl}
    \gsep\ \xcomp{\xl}{\xl}
    \gsep\ \xlvar\
    \gsep\ \xlab{\xlvar}{\xl}{\xl}

%% file: definitions/MRLens_DenotationalSemantics.tex
\begin{align*}
\denotes{\xconst{\s[1]}{\s[2]}} 
    &= \set{\xpair{\s[1]}{\s[2]}} \\
\denotes{\xid{\mathR}} 
    &= \defset{\xpair{\s}{\s}}{\s \xin \rL[\empenv]{\mathR}}\\
\denotes{\xiter{\xl}} 
    &= \defset{\xpair{\rand{\rand{\s[11]}{\dots}}{\s[1n]}}{\rand{\rand{\s[21]}{\dots}}{\s[2n]}}}
              {\xpair{\s[1k]}{\s[2k]} \xin \denotes{\xl}} \\
\denotes{\xconcat{\xl[1]}{\xl[2]}} 
    &= \defset{\xpair{\rand{\s[11]}{\s[12]}}{\rand{\s[21]}{\s[22]}}}
              {\xpair{\s[11]}{\s[21]} \xin \denotes{\xl[1]} 
                \xand \xpair{\s[12]}{\s[22]} \xin \denotes{\xl[2]}} \\
\denotes{\xswap{\xl[1]}{\xl[2]}} 
    &= \defset{\xpair{\rand{\s[11]}{\s[12]}}{\rand{\s[22]}{\s[21]}}}
              {\xpair{\s[11]}{\s[21]} \xin \denotes{\xl[1]} 
                              \xand \xpair{\s[12]}{\s[22]} \xin \denotes{\xl[2]}}  \\
\denotes{\xlor{\xl[1]}{\xl[2]}} 
    &= \defset{\xpair{\s[1]}{\s[2]}}
              {\xpair{\s[1]}{\s[2]} \xin \denotes{\xl[1]} \vee \xpair{\s[1]}{\s[2]} \xin \denotes{\xl[2]}} \\
\denotes{\xcomp{\xl[1]}{\xl[2]}} 
    &= \defset{\xpair{\s[1]}{\s[3]}}
              {\xpair{\s[1]}{\s[2]} \xin \denotes{\xl[1]} \xand \xpair{\s[2]}{\s[3]} \xin \denotes{\xl[2]}} \\
\denotes{\xlvar} 
    &= \set{\lu{\lenv}{\xlvar}}\\
\denotes{\xlab{\xlvar}{\xl[1]}{\xl[2]}} 
    &= \defset{(\s[1], \s[2])}
              {(\xpr{\s[1]}, \xpr{\s[2]}) \xin \denotes{\xl[1]} 
                \xand \xpair{\s[1]}{\s[2]} \xin \denotes[\extenv{\lenv}{\xlvar}{(\xpr{\s[1]}, \xpr{\s[2]})}]{\xl[2]}} 
\end{align*}

%% file: definitions/MRLens_TypingRules.tex
\begin{mathpar}           
\inferrule*[lab=ConstT]
          {\goodtyenvs \and \s[1], \s[2] \xin Strings}
          {\ljudg{\xconst{\s[1]}{\s[2]} \ANNOT{\s[1]}{\s[2]}}
                 {\ltype{\s[1]}{\s[2]}}}

\inferrule*[lab=VarT]
    {\goodtyenvs
     \and \lu{\ltenv}{\xlvar} = \ltype{\dvarx[1]}{\dvarx[2]}}
    {\ljudg{\xlvar \ANNOT{\dvarx[1]}{\dvarx[2]}}
    {\ltype{\dvarx[1]}{\dvarx[2]}}}  
              
\inferrule*[left=IdT]
          {\goodtyenvs 
            \and \mathR \text{ is strongly unambiguous in } \empenv 
            \and \mathR \text{ is closed}}
          {\ljudg{\xid{\mathR} \ANNOT{\mathR}{\mathR}}
                 {\ltype{\mathR}{\mathR}}}
\\\\
\inferrule*[lab=IterT]
            {\ljudg{\xl \ANNOT{\mathR[1]}{\mathR[2]}}
                  {\ltype{\mathR[1]}{\mathR[2]}} 
              \\\\ \uiter{\mathR[1]} \and \uiter{\mathR[2]}}
            {\ljudg{\xiter{\xl}
                \ANNOT{\rstar{\mathR[1]}}{\rstar{\mathR[2]}}}
            {\ltype{\rstar{\mathR[1]}}{\rstar{\mathR[2]}}}}

\inferrule*[lab=OrT]
          {\ljudg{\xl[1]\ANNOT{\mathR[11]}{\mathR[21]}}{\ltype{\mathR[11]}{\mathR[21]}}
          \\\\ \ljudg{\xl[2]\ANNOT{\mathR[12]}{\mathR[22]}}
                     {\ltype{\mathR[12]}{\mathR[22]}}                     
          \\\\ \ualt{\mathR[11]}{\mathR[12]} \and \ualt{\mathR[21]}{\mathR[22]}}
          {\ljudg{\xlor{\xl[1]}{\xl[2]} \ANNOT{\ralt{\mathR[11]}{\mathR[12]}}{\ralt{\mathR[21]}{\mathR[22]}}}
                 {\ltype{\ralt{\mathR[11]}{\mathR[12]}}{\ralt{\mathR[21]}{\mathR[22]}}}}
                           
\inferrule*[lab=ConcatT]
          {\ljudg{\xl[1] \ANNOT{\mathR[11]}{\mathR[21]}}
                 {\ltype{\mathR[11]}{\mathR[21]}}
          \\\\ \ljudg{\xl[2]\ANNOT{\mathR[12]}{\mathR[22]}}
                      {\ltype{\mathR[12]}{\mathR[22]}}
          \\\\ \uconcat{\mathR[11]}{\mathR[12]} \and \uconcat{\mathR[21]}{\mathR[22]}}
          {\ljudg{\xconcat{\xl[1]}{\xl[2]}
                  \ANNOT{\rand{\mathR[11]}{\mathR[12]}}
                        {\rand{\mathR[21]}{\mathR[22]}}}
                 {\ltype{\rand{\mathR[11]}{\mathR[12]}}{\rand{\mathR[21]}{\mathR[22]}}}}
                 
\inferrule*[lab=SwapT]
          {\ljudg{\xl[1] \ANNOT{\mathR[11]}{\mathR[21]}}
                 {\ltype{\mathR[11]}{\mathR[21]}}
          \\\\ \ljudg{\xl[2] \ANNOT{\mathR[12]}{\mathR[22]}}
                      {\ltype{\mathR[12]}{\mathR[22]}}
          \\\\ \uconcat{\mathR[11]}{\mathR[12]} \and \uconcat{\mathR[22]}{\mathR[21]}}
          {\ljudg{\xswap{\xl[1]}{\xl[2]} \ANNOT{\rand{\mathR[11]}{\mathR[12]}}{\rand{\mathR[22]}{\mathR[21]}}}
                 {\ltype{\rand{\mathR[11]}{\mathR[12]}}{\rand{\mathR[22]}{\mathR[21]}}}}
                 
\inferrule*[left=CompT]
          {\ljudg{\xl[1] \ANNOT{\mathR[1]}{\mathR[2]}}
                 {\ltype{\mathR[1]}{\mathR[2]}}
          \\\\ 
          \ljudg{\xl[2] \ANNOT{\mathR[2]}{\mathR[3]}}
                 {\ltype{\mathR[2]}{\mathR[3]}}
          }
          {\ljudg{\xcomp{\xl[1]}{\xl[2]} \ANNOT{\mathR[1]}{\mathR[3]}}
                 {\ltype{\mathR[1]}{\mathR[3]}}}
                 
\inferrule*[left=LinkT]
        {\ljudg{\xl[1] \ANNOT{\mathR[11]}{\mathR[21]}}
               {\ltype{\mathR[11]}{\mathR[21]}}
        \\\\ 
        \dvarx[1], \dvarx[2] \text{ are fresh variables}
        \\\\ 
        \ljudg[\exte{\ltenv}{\mape{\xlvar}{\ltype{\dvarx[1]}{\dvarx[2]}}},
            \exte{\rtenv}{\mape{\dvarx[1]}{\mathR[11]}, \mape{\dvarx[2]}{\mathR[21]}}]
           {\xl[2] \ANNOT{\mathR[12]}{\mathR[22]}}{\ltype{\mathR[12]}{\mathR[22]}}
        \\\\ 
        \ubind{\dvarx[1]}{\mathR[11]}{\mathR[12]}
        \and \ubind{\dvarx[2]}{\mathR[21]}{\mathR[22]}}
        {\ljudg{\xlab{\xlvar}{\xl[1]}{\xl[2]}     
            \ANNOT{\bindin{\dvarx[1]}{\mathR[11]}{\mathR[12]}}
                  {\bindin{\dvarx[2]}{\mathR[21]}{\mathR[22]}}}
       {\ltype{\bindin{\dvarx[1]}{\mathR[11]}{\mathR[12]}}
              {\bindin{\dvarx[2]}{\mathR[21]}{\mathR[22]}}}}
\end{mathpar} 

%% file: definitions/MRLens_BigStepSemantics.tex
\begin{mathpar}
\begin{minipage}[]{.4\textwidth}
\begin{framed}
        $f \xin \{ \lget, \lput \}$
        \\
        $f = \lget \xthen j = 1, k = 2$
        \\
        $f = \lput \xthen j = 2, k = 1$
    \end{framed}
\end{minipage}

    \inferrule*[lab=EStep]
        {
            e \Downarrow \s[j] \and \fstep{\xl}{\s[j]}{\s[k]}}
        {\fstep{\xl}{e}{\s[k]}}
        
     \inferrule*[lab=ConstG/P]
         {\ltenv, \rtenv, \renv \vdash * }
         {\fstep{\xconst{\s[1]}{\s[2]}}{\s[j]}{\s[k]}}

    \inferrule*[lab=VarG/P]
        {\goodstepenvs
              	\and \ljudg{\xlvar \ANNOT{\dvarx[1]}{\dvarx[2]}}{\ltype{\dvarx[1]}{\dvarx[2]}}
            \and \s[j] \xin \rL{\dvarx[j]}}
        {\fstep{\xlvar}{\s[j]}{\lu{\renv}{\dvarx[k]}}}
              
     \inferrule*[left=IdG/P]
         {\ltenv, \rtenv, \renv \vdash * 
            \and \ljudg{\xid{\mathR} \ANNOT{\mathR}{\mathR}}{\ltype{\mathR}{\mathR}}
            \and \s \xin \rL[\empenv]{\mathR}}
         {\fstep{\xid{\mathR}}{\s}{\s}}       
         
    \inferrule*[left=IterG/P]
        {\ljudg{\xiter{\xl} \ANNOT{\rstar{\mathR[1]}}{\rstar{\mathR[2]}}}
            {\ltype{\rstar{\mathR[1]}}{\rstar{\mathR[2]}}}
         \\\\ \s[j] = \rand{\rand{\s[j1]}{...}}{\s[jn]}
1         \\ \s[j1] \xin \rL{\mathR[j]}, \ ... \ , \s[jn] \xin \rL{\mathR[j]}
         \\\\ \fstep{\xl}{\s[j1]}{\s[k1]}, \ ... \ , \fstep{\xl}{\s[jn]}{\s[kn]}
         \and  \s[k] = \rand{\rand{\s[k1]}{...}}{\s[kn]}}
        {\fstep{\xiter{\xl}}{\s[j]}{\s[k]}} 
        
    \inferrule*[left=ConcG/P]
        {\ljudg{\xconcat{\xl[1]}{\xl[2]} 
                \ANNOT{\rand{\mathR[11]}{\mathR[12]}}{\rand{\mathR[21]}{\mathR[22]}}}
              {\ltype{\rand{\mathR[11]}{\mathR[12]}}
                      {\rand{\mathR[21]}{\mathR[22]}}}                     
         \\\\ \s[j] = \rand{\s[j1]}{\s[j2]} 
         \\ \s[j1] \xin \rL{\mathR[j1]}
         \and \s[j2] \xin \rL{\mathR[j2]}
         \\\\ \fstep{\xl[1]}{\s[j1]}{\s[k1]}
         \and \fstep{\xl[2]}{\s[j2]}{\s[k2]}
         \and \s[k] = \rand{\s[k1]}{\s[k2]}}
        {\fstep{\xconcat{\xl[1]}{\xl[2]}}{\s[j]}{\s[k]}} 
    
    \inferrule*[left=SwapG/P]
        {\ljudg{\xswap{\xl[1]}{\xl[2]} 
            \ANNOT{\rand{\mathR[11]}{\mathR[12]}}
                  {\rand{\mathR[22]}{\mathR[21]}}}
              {\ltype{\rand{\mathR[11]}{\mathR[12]}}
                      {\rand{\mathR[22]}{\mathR[21]}}}
            \\\\ \s[j] = \rand{\s[j1]}{\s[j2]}
            \\ \s[j1] \xin \rL{\mathR[j1]}
            \and \s[j2] \xin \rL{\mathR[j2]}
            \\\\ \fstep{\xl[1]}{\s[j1]}{\s[k1]}
            \and \fstep{\xl[2]}{\s[j2]}{\s[k2]}
            \and \s[k] = \rand{\s[k2]}{\s[k1]}}
        {\fstep{\xswap{\xl[1]}{\xl[2]}}{\s[j]}{\s[k]}}

    \inferrule*[lab=OrLG/P]
        {\ljudg{\xlor{\xl[1]}{\xl[2]}
                \ANNOT{\ralt{\mathR[11]}{\mathR[12]}}{\ralt{\mathR[21]}{\mathR[22]}}}
              {\ltype{\ralt{\mathR[11]}{\mathR[12]}}
                      {\ralt{\mathR[21]}{\mathR[22]}}}
        \\\\
        \s[j] \xin \rL{\mathR[j1]}
            \and \fstep{\xl[1]}{\s[j]}{\s[k]}
        }
        {\fstep{\xlor{\xl[1]}{\xl[2]}}{\s[j]}{\s[k]}}   
               
    \inferrule*[lab=OrRG/P]
        {\ljudg{\xlor{\xl[1]}{\xl[2]}
                \ANNOT{\ralt{\mathR[11]}{\mathR[12]}}{\ralt{\mathR[21]}{\mathR[22]}}}
              {\ltype{\ralt{\mathR[11]}{\mathR[12]}}
                      {\ralt{\mathR[21]}{\mathR[22]}}}
        \\\\  \s[j] \xin \rL{\mathR[j2]}
            \and \fstep{\xl[2]}{\s[j]}{\s[k]}
        }
        {\fstep{\xlor{\xl[1]}{\xl[2]}}{\s[j]}{\s[k]}}

     \inferrule*[left=CompG/P]
         {\ljudg{\xcomp{\xl[1]}{\xl[2]} \ANNOT{\mathR[1]}{\mathR[2]}}
         {\ltype{\mathR[1]}{\mathR[2]}}
          \\\\ \s[j] \xin \rL{\mathR[j]}
          \and \fstep{\xl[j]}{\s[j]}{\s[i]}
          \and \fstep{\xl[k]}{\s[i]}{\s[k]}
          }
         {\fstep{\xcomp{\xl[1]}{\xl[2]}}{\s[j]}{\s[k]}}
         
\inferrule*[left=LinkG/P]
        {\ljudg{\xlab{\xlvar}{\xl[1]}{\xl[2]}
                \ANNOT{\bindin{\dvarx[1]}{\mathR[11]}{\mathR[12]}}{\bindin{\dvarx[2]}{\mathR[21]}{\mathR[22]}}}
               {\ltype{\bindin{\dvarx[1]}{\mathR[11]}{\mathR[12]}}
                      {\bindin{\dvarx[2]}{\mathR[21]}{\mathR[22]}}}
          \\\\ \s[j1] \xin \rL{\mathR[j1]}
          \and \s[j2] \xin \rL[\exte{\renv}{\mape{\dvarx[j]}{\s[j1]}}]{\mathR[j2]} 
          \\\\ \fstep{\xl[1]}{\s[j1]}{\s[k1]}
          \\\\ \fstep[\exte{\ltenv}{\mape{\xlvar}{\ltype{\dvarx[1]}{\dvarx[2]}}},
                				\exte{\rtenv}{\mape{\dvarx[1]}{\mathR[11]}, \mape{\dvarx[2]}{\mathR[21]}},
                        \exte{\renv}{\mape{\dvarx[1]}{\s[11]}, \mape{\dvarx[2]}{\s[21]}}]
                       {\xl[2]}{\s[j2]}{\s[k2]}}
        {\fstep{\xlab{\xlvar}{\xl[1]}{\xl[2]}}{\s[j2]}{\s[k2]}}
\end{mathpar}

%% file: definitions/MRAS_steps.tex
    \textbf{Consume Input} 
        \begin{mathpar}
            \inferrule[\lcmatch]
                {(\rfsaq{i}{k}, \xpref, \rfsaq{i}{k'}) \xin \getfsad{\lu{\rfsaset}{i}}
                \\\\ \xpref \xin \rasalph \xunion \set{\lambda}
                \\\\ 
                \lu{\metavbuf}{i} = \vbcurr{\xstr'} \and \lu{\metavbuf'}{i} = \vbcurr{\xstr'\xpref} 
                \\\\
                \xfa j \neq i, \lu{\metavbuf}{j} = \lu{\metavbuf'}{j}}
                {(\rfsaq{i}{k}, \xpref\xstr, \rasstack, \metavbuf) \rstep (\rfsaq{i}{k'}, \xstr, \rasstack, \metavbuf')}  
    
            \inferrule[\lvmatch]
                {(\rfsaq{i}{k}, \rasvar{j}, \rfsaq{i}{k'}) \xin \getfsad{\lu{\rfsaset}{i}}
                \\\\ \lu{\metavbuf}{j} = \vbfound{\xstr'} 
                \\\\
                \lu{\metavbuf}{i} = \vbcurr{\xstr''} \and \lu{\metavbuf'}{i} = \vbcurr{\xstr''\xstr'} 
                \\\\
                \xfa j \neq i, \lu{\metavbuf}{j} = \lu{\metavbuf'}{j}}
                {(\rfsaq{i}{k}, \xstr'\xstr, \rasstack, \metavbuf) \rstep (\rfsaq{i}{k'}, \xstr, \rasstack, \metavbuf')}
        \end{mathpar}

    \textbf{Update Variable Scope}
        \begin{mathpar}
            \inferrule[\lvin]
                {(\rfsaq{i}{k}, \rasinvar{j}, \rfsaq{i}{k'}) \xin \getfsad{\lu{\rfsaset}{i}} 
                \\\\ 
                \lu{\metavbuf}{j} = \vbout \and \lu{\metavbuf'}{j} = \vbin  
                \\\\
                \xfa h \neq j, \lu{\metavbuf}{h} = \lu{\metavbuf'}{h}}
                {(\rfsaq{i}{k}, \xstr, \rasstack, \metavbuf) \rstep (\rfsaq{i}{k'}, \xstr, \rasstack, \metavbuf')}   
                
            \inferrule[\lvout]
                {(\rfsaq{i}{k}, \rasoutvar{j}, \rfsaq{i}{k'}) \xin \getfsad{\lu{\rfsaset}{i}} 
                \\\\ 
                \lu{\metavbuf}{j} = \vbin \vee \vbfound{\xpr{\xstr}} \and \lu{\metavbuf'}{j} = \vbout  
                \\\\
                \xfa h \neq j, \lu{\metavbuf}{h} = \lu{\metavbuf'}{h}}
                {(\rfsaq{i}{k}, \xstr, \rasstack, \metavbuf) \rstep (\rfsaq{i}{k'}, \xstr, \rasstack, \metavbuf')}
        \end{mathpar}

    \textbf{Switch Automaton}
        \begin{mathpar}        
            \inferrule[\linitv]
                {(\rfsaq{i}{k}, \rasvar{j}, \rfsaq{i}{k'}) \xin \getfsad{\lu{\rfsaset}{i}} 
                \\\\ 
                 \rasstack' = \stpush{\rasstack}{\rfsaq{i}{k'}}
                \\\\
                \lu{\metavbuf}{j} = \vbin  \and \lu{\metavbuf'}{j} = \vbcurr{\xemp}  
                \\\\
                \xfa h \neq j, \lu{\metavbuf}{h} = \lu{\metavbuf'}{h}}
                {(\rfsaq{i}{k}, \xstr, \rasstack, \metavbuf) \rstep (\rfsainit{j}, \xstr, \rasstack', \metavbuf')}
         
            \inferrule[\lvret]
                {\rfsaq{i}{k} \xin  \getfsaqf{\lu{\rfsaset}{i}} 
                \\\\ \rfsaq{i'}{k'} = 
                \sttop{\rasstack}
                 \\\\            
                 \rasstack' = \stpop{\rasstack}
                \\\\
                \lu{\metavbuf}{i} = \vbcurr{\xstr'} \and \lu{\metavbuf'}{i} = \vbfound{\xstr'}
                \\\\
                \lu{\metavbuf}{i'} = \vbcurr{\xstr''} \and \lu{\metavbuf'}{i'} = \vbcurr{\xstr''\xstr'}
                \\\\
                \xfa h \neq i,i', \lu{\metavbuf}{h} = \lu{\metavbuf'}{h}
                }
                {(\rfsaq{i}{k}, \xstr, \rasstack, \metavbuf) \rstep (\rfsaq{i'}{k'}, \xstr, \xpr{\rasstack}, \metavbuf')} 
        \end{mathpar}

%% file: definitions/SR_denotational_simple.tex
    \begin{align*}
        \rL{\sre{\rconst{c}}{\sd}} 
            &= \set{\rconst{c}}\\
        \rL{\sre{\dvarx}{\sd}} 
            &= \set{\lu{\renv}{\dvarx}} \\
        \rL{\sre{\rstar{\rpart}}{\sd}} 
            &= \rstar{\rL{\sre{\rpart}{\sd}}}\\
        \rL{\sre{\ralt{\rpart[1]}{\rpart[2]}}{\sd}} 
            &= \rL{\sre{\rpart[1]}{\sd}} \xunion \rL{\sre{\rpart[2]}{\sd}} \\
        \rL{\sre{\rand{\rpart[1]}{\rpart[2]}}{\sd}} 
            &= \rand{\rL{\sre{\rpart[1]}{\sd}}}{\rL{\sre{\rpart[2]}{\sd}}} \\
        \rL{\sre{\rscope{\rpart}{\dvarx}}{\sd}} 
            &=  \defset{\s}
                                  {\s' \xin \rL{\sre{\lu{\sd}{\dvarx}}{\subd{\sd}{\dvarx}}} 
                                   \xand \s \xin \rL[\extenv{\renv}{\dvarx}{\s'}]{\sre{\rpart}{\sd}}}
    \end{align*}

%% file: definitions/rpart-to-fsa.tex
\begin{tabular}{|c||l|l|l|l|}
    \multicolumn{5}{l}{\rtofsa{\rpart}{i} = \rfsacomp[\rpart]{i} \hfill \rasalph[i] = \rasalph\ \xunion \defset{sym}{0 
        \leq k < i \xand sym \xin \set{\rasvar{k}, \rasinvar{k}, \rasoutvar{k}}}} 
    \\ \hline
    \rpart\ 
        & \multicolumn{1}{c|}{\emath{Q_r}} 
        & \multicolumn{1}{c|}{\emath{\delta_r}} 
        & \multicolumn{1}{c|}{\emath{q_{I_r}}} 
        & \multicolumn{1}{c|}{\emath{Q_{F_r}}}
        \\ \hline
    \rconst{c} 
        & {\set{\rfsaq{i}{j^{\dag}}, \rfsaq{i}{k^{\dag}}}}
        & {\set{\fd{\rfsaq{i}{j^{\dag}}}{\rconst{c}}{\rfsaq{i}{k^{\dag}}}}}
        & {\rfsaq{i}{j^{\dag}}} 
        & {\set{\rfsaq{i}{k^{\dag}}}} 
        \\ \hline
    \dvarx[h] 
        & {\set{\rfsaq{i}{j^{\dag}}, \rfsaq{i}{k^{\dag}}}}
        & {\set{\fd{\rfsaq{i}{j^{\dag}}}{\rasvar{h}}{\rfsaq{i}{k^{\dag}}}}}
        & {\rfsaq{i}{j^{\dag}}} 
        & {\set{\rfsaq{i}{k^{\dag}}}} 
        \\ \hline
    \rstar{\xpr{\rpart}}
        & {\rfsaqset{\xpr{\rpart}} 
                        \xunion \set{\rfsaq{i}{j^{\dag}}} }
       & {\rfsastep{\xpr{\rpart}} \xunion \set{\fd{\rfsaq{i}{j^{\dag}}}{\varepsilon}{q_{I_{\xpr{\rpart}}}}}
                   \xunion \defset{\fd{q}{\varepsilon}{\rfsainit{\xpr{\rpart}}}}{q \xin \rfsafinal{\xpr{\rpart}}}}
       & {\rfsaq{i}{j^{\dag}}}
       & {\rfsafinal{\xpr{\rpart}} \xunion \set{\rfsaq{i}{j^{\dag}}}} 
       \\ \hline  
    \ralt{\rpart[1]}{\rpart[2]}
        & {\rfsaqset{\rpart[1]} \xunion \rfsaqset{\rpart[2]} \xunion \set{\rfsaq{i}{j^{\dag}}}}
        & {\rfsastep{\rpart[1]} \xunion \rfsastep{\rpart[2]} 
             \xunion \set{\fd{\rfsaq{i}{j^{\dag}}}{\varepsilon}{\rfsainit{\rpart[1]}}, \fd{\rfsaq{i}{j^{\dag}}}{\varepsilon}{q_{I_{\rpart[2]}}}}}
        & {\rfsaq{i}{j^{\dag}}}
        &  {\rfsafinal{\rpart[1]} \xunion  \rfsafinal{\rpart[2]}}
        \\ \hline
    \rand{\rpart[1]}{\rpart[2]}
        & {\rfsaqset{\rpart[1]} \xunion \rfsaqset{\rpart[2]}}
        & {\rfsastep{\rpart[1]} \xunion \rfsastep{\rpart[2]} 
            \xunion \defset{\fd{q}{\varepsilon}{\rfsainit{\rpart[2]}}}{q \xin \rfsafinal{\rpart[1]}}}
        & {\rfsainit{\rpart[1]}}
        & {\rfsafinal{\rpart[2]}} 
        \\ \hline
    \rscope{\xpr{\rpart}}{\dvarx[h]}
        & {\rfsaqset{\xpr{\rpart}} \xunion \set{\rfsaq{i}{j^{\dag}}, \rfsaq{i}{k^{\dag}}}}
        & {\rfsastep{\xpr{\rpart}} \xunion  \set{\fd{\rfsaq{i}{j^{\dag}}}{\rasinvar{h}}{\rfsainit{\xpr{\rpart}}}}
               \xunion \defset{\fd{q}{\rasoutvar{h}}{\rfsaq{i}{k^{\dag}}}}{q \xin \rfsafinal{\xpr{\rpart}}}}
        & {\rfsaq{i}{j^{\dag}}}
        & {\set{\rfsaq{i}{k^{\dag}}}} \\ \hline
    \multicolumn{5}{r}{\textit{note}: the \emath{\dag} in \emath{\rfsaq{i}{k^\dag}} indicates that this is a fresh state} \\  
\end{tabular}